\documentclass[12pt]{amsart}
\usepackage[round]{natbib}
\usepackage{times}
\usepackage{amsmath}
\usepackage{amssymb}
\usepackage{amsfonts}
\usepackage{graphics}
\usepackage{subfigure}
\usepackage{framed}
\usepackage{graphicx}
\usepackage[dvipsnames]{color}

\usepackage{sidecap}
\usepackage{wrapfig}
\usepackage{subfig}

\usepackage{url}

\usepackage[left=2cm,top=2cm,bottom=2cm,right=2cm,nofoot]{geometry} 
\definecolor{shadecolor}{gray}{0.85}

\newtheorem{theorem}{Theorem}

\newcommand{\calE}{\mathcal{E}}

\newcommand{\specialbox}[2]{
\begin{shaded}
\centerline{\large\textsc{#1}}\medskip%
\nopagebreak
\small #2
\end{shaded}
}

\title{Computational Tools for Evaluating Phylogenetic   and Hierarchical Clustering Trees}
\author{John Chakerian and Susan Holmes}
\thanks{Research funded by NIH grant R01GM086884 and NSF grant DMS-0241246}
\address{Statistics Department-Sequoia Hall\\Stanford CA94305}
\email{susan@stat.stanford.edu}
\begin{document}
\maketitle
\begin{abstract}
  Inferential summaries of tree estimates are useful in the setting of
  evolutionary biology, where phylogenetic trees have been built from DNA
  data since the 1960's. In bioinformatics, psychometrics and data mining,
  hierarchical clustering techniques output the same mathematical objects,
  and practitioners have similar questions about the stability and
  `generalizability' of these summaries.  This paper provides an
  implementation of the geometric distance between trees developed by
  \citet*{Bhv} equally applicable to phylogenetic trees and heirarchical
  clustering trees, and shows some of the applications in statistical
  inference for which this distance can be useful.

  In particular, since \citet{Bhv} have shown that the space of trees is
  negatively curved (a {\bf CAT(0)} space), a natural representation of a
  collection of trees is a tree. We compare this representation to the
  Euclidean approximations of treespace made available through
  Multidimensional Scaling of the matrix of distances between trees. We
  also provide applications of the distances between trees to hierarchical
  clustering trees constructed from microarrays. Our method gives a new
  way of evaluating the influence both of certain columns (positions,
  variables or genes) and of certain rows (whether species, observations
  or arrays).
\end{abstract}

\section{Current Practices in Estimation and Stability of Hierarchical Trees}

Trees are often used as a parameter in phylogenetic studies and for data
description in hierarchical clustering and regression through procedures
like CART   \citep{cart}.  These methods can generate many trees leading to
the need for summaries and methods of analysis and display.

\begin{figure}[t] 
\begin{minipage}{3.2in}
  \includegraphics[width=2.9in]{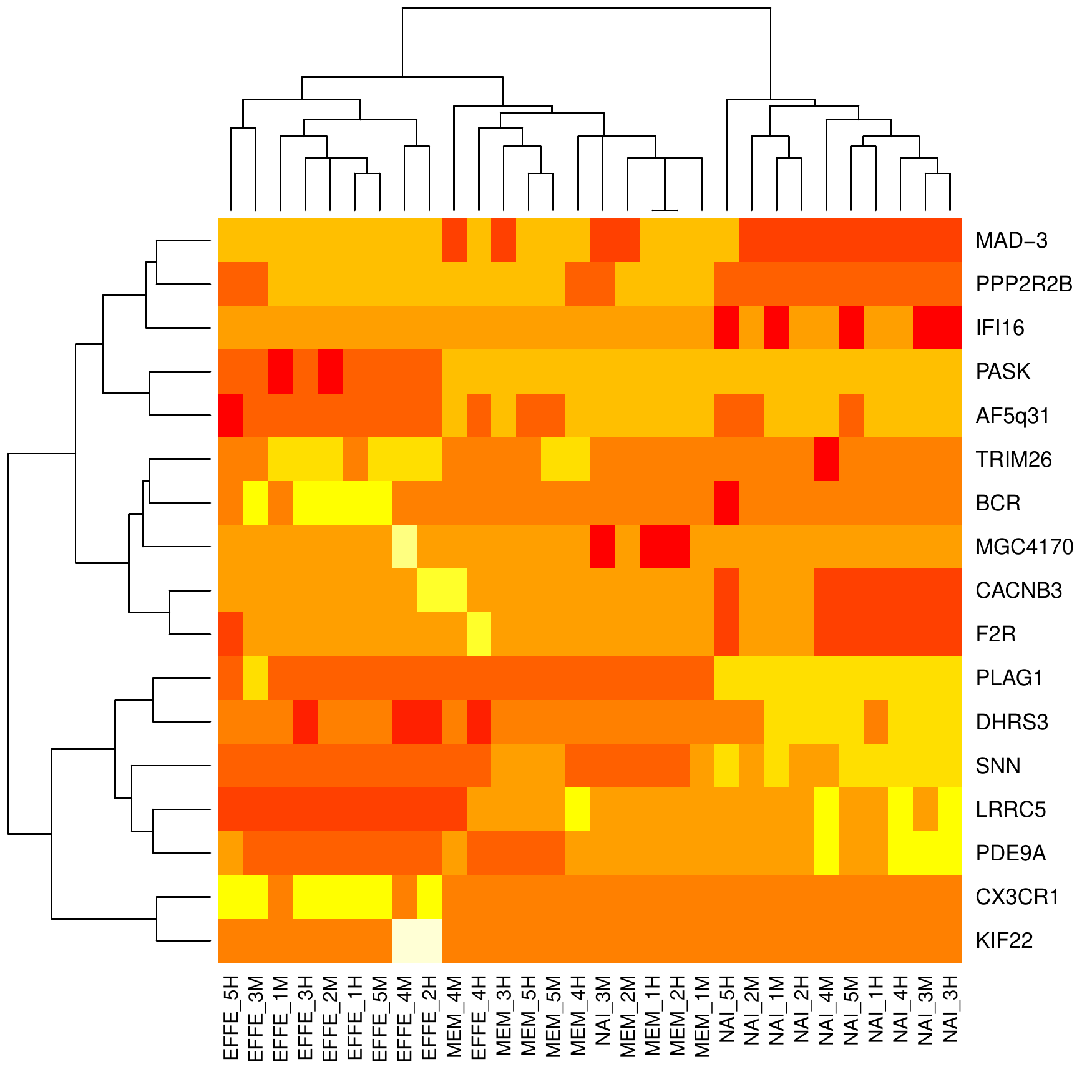} \\
  { (a) Hierarchical Clustering trees of both rows and columns of a
    microarray matrix. Rows are genes, columns are patients.}
  \end{minipage}
  \hskip10pt
  \begin{minipage}{3.2in}
      \includegraphics[width=2.9in]{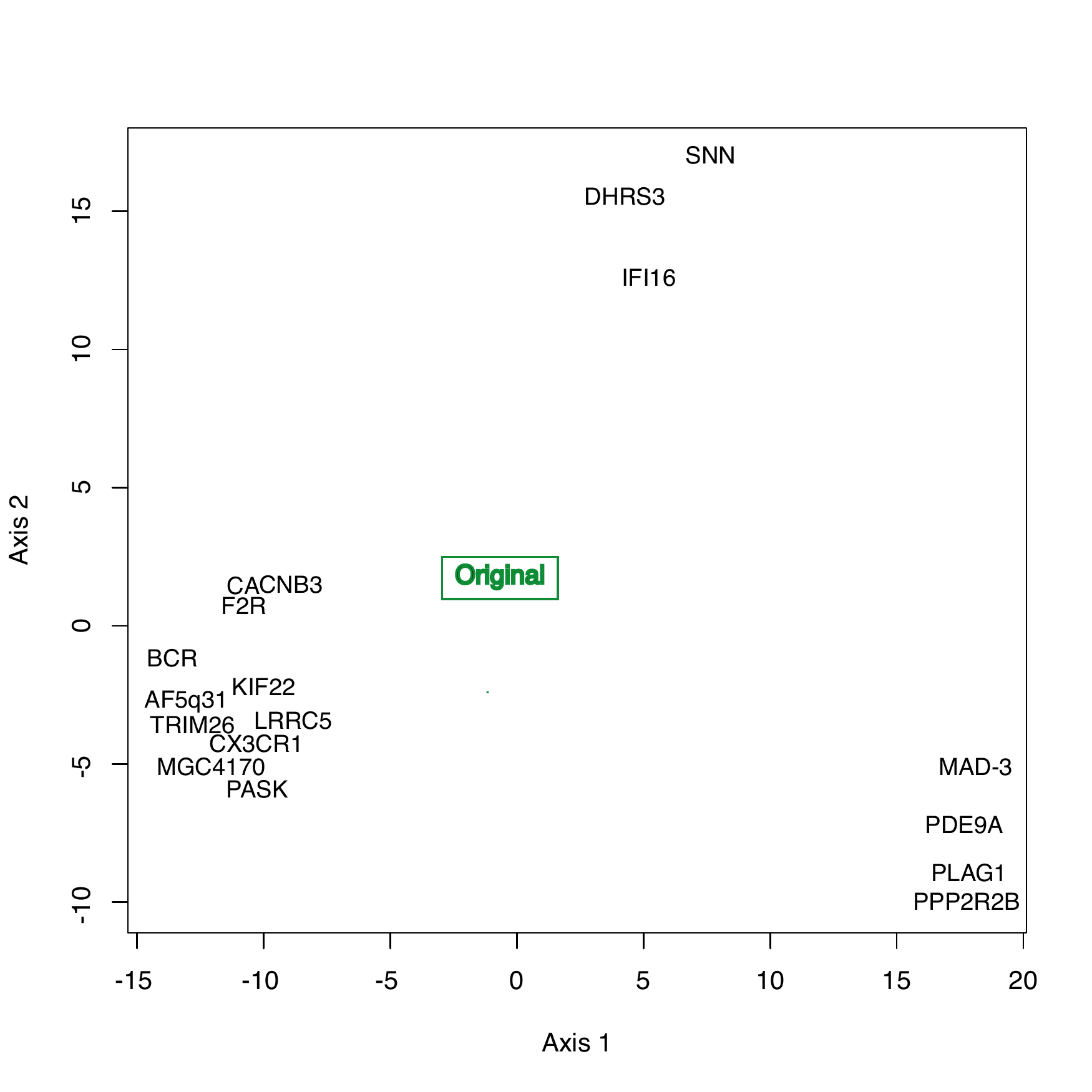} \\
      { (b) Plot of cross validated hierarchical clustering trees. Each
        point represents a tree that was estimated without the gene it is
        labeled with.}
   \end{minipage}
   \caption{Cross validation of the rows allows us to geometrically study
     the leverage of individual genes.}
   \label{fig:hclustcv}
\end{figure}

A natural distance between trees has been introduced and studied in
\citet*{Bhv}. Recent advances in its computation   \citep{owenprovan}, along
with advances reported below, now allow for efficient computation and use
of this distance. The present paper describes these advances and some
applications. We begin with a small example.
\\
\vskip5pt
{\bf Example 1: Hierarchical Clustering variability}\\
Now a staple of microarray visualizations, the hierarchical clustering
trees such as that in Figure~ \ref{fig:hclustcv} is a standard {\em
  heatmap} type plot showing both a clustering of patients (the columns in
this data) and the genes (the rows).

We will consider using cross validation to see how the clustering trees
change when each of the genes was removed. This gives us 16 cross
validated trees and the original tree. These 17 points can be represented
in a plane, where the groupings show that some genes have similar effects
on the estimates when missing. Figure ~\ref{fig:hclustcv} shows the cross
validated trees with the original tree at the center of the triangular
scatter of points.  Notice that the cross validated trees can be seen to
form three clusters.  We will explain how this plot was made in section
3.6.

\subsection{Trees as Statistical Summaries}
Hierarchical clustering trees and phylogenetic trees are some of the most
popular graphical representations in contemporary evolutionary biology and
data analysis.  They share a common non-standard output: a binary rooted
tree with the known entities at the leaves.  Mathematicians often call
them rooted semi-labeled binary trees  \citep{semplesteel}.  The trees are
built from multivariate data sets with data on the leaves; we will suppose
the data are organized so that each row corresponds to a leaf.
 
Current practices in evaluating tree estimates lean on unidimensional
summaries giving the proportion of times a clade occurs. These are
recorded either as the binomial success rates along the branches of the
tree  \citep{felsen} or as a set of bin frequencies of the competing trees
considered as categorical output.

In this paper we propose alternative evaluation procedures, all based on
distances between trees.  The idea of comparing trees through a notion of
distance between trees has many variations.  \cite{RobinsonF} proposed a
coarse distance between phylogenetic trees that takes on only integer
values; \cite{waterman} proposed the Nearest Neighbor Interchange (NNI) as
a biologically reasonable distance between trees. We will use that of
\citet*{Bhv} and call it the {\bf BHV} distance. This distance can be
considered in some sense to be a refinement of NNI that comes from taking
into account the edge lengths  of the trees.

In this first section we will place the question of evaluating trees in
the context of statistical estimation and give an overview of current
practices in data analysis.  Here our data will be presented as a $n\times
p$ matrix, with n being the number of observations for the hierarcical
clustering studies or the number of species for the phylogenetic examples.
The elements of the matrix will mostly come from small alphabets; examples
in the paper include ${\mathcal A}= \{ 0,1\},\mbox{ or }\{-1,0,1\}\mbox{
  or } \{a,c,g,t\}$.
 
The second section will provide a short description of the algorithm for
computing the distances between trees as we have implemented it.  Section
3 shows how we can use multidimensional scaling to approximately embed the
trees in a Euclidean space. We show examples of using multidimensional
representations for comparing trees generated from different data, and for
comparing cross validated data for detecting influential variables in
hierarchical clustering. Section 4 shows how we embed the trees in a tree,
providing a robust method for detecting mixtures. We also introduce a
quantitative measure of treeness that  tells us how appropriate a tree
representation might be.
 
Section 5 shows how paths between trees can be used to find the boundary
points between two different branching orders. These paths are built using
simulated annealing algorithm and can also provide the boundary data used
by \cite*{efronh} to correct the bias in the na\"ive bootstrap for trees
  \citep{Holmes-2003}.

\subsection{Estimating phylogenetic trees from Data}

The true tree, if one exists, can be considered an unknown parameter that
we can use standard statistical estimation methods to estimate
  \citep{holmesima}.  Recently, it has become apparent that in many cases
there are probably several good candidate trees that must be presented
together to explain the complexities of the evolutionary process. Having
several trees to represent increases the need for a satisfactory
comparison technique. We will begin, however, with the simplest case, that
of having only one true tree with a simple model of evolution.

If the data are the result of a simple treelike evolutionary process, we
may model the process as a Markov chain. We can characterize it by a pair
of parameters $(\tau, M)$, where $\tau$ represents the tree with its edge
lengths and $M$ is the mutation (transition) matrix. If the characters
measured at the leaves of the tree are binary, M will be a $2\times2$
transition matrix; in the familiar case of observed DNA, $M$ will be a
$4\times4$ transition matrix.

\subsection*{Example 2: Balanced phylogenetic tree and comb-like phylogenetic trees. }
\begin{figure}[h!]
   \includegraphics[height=3.2in,width=3in]{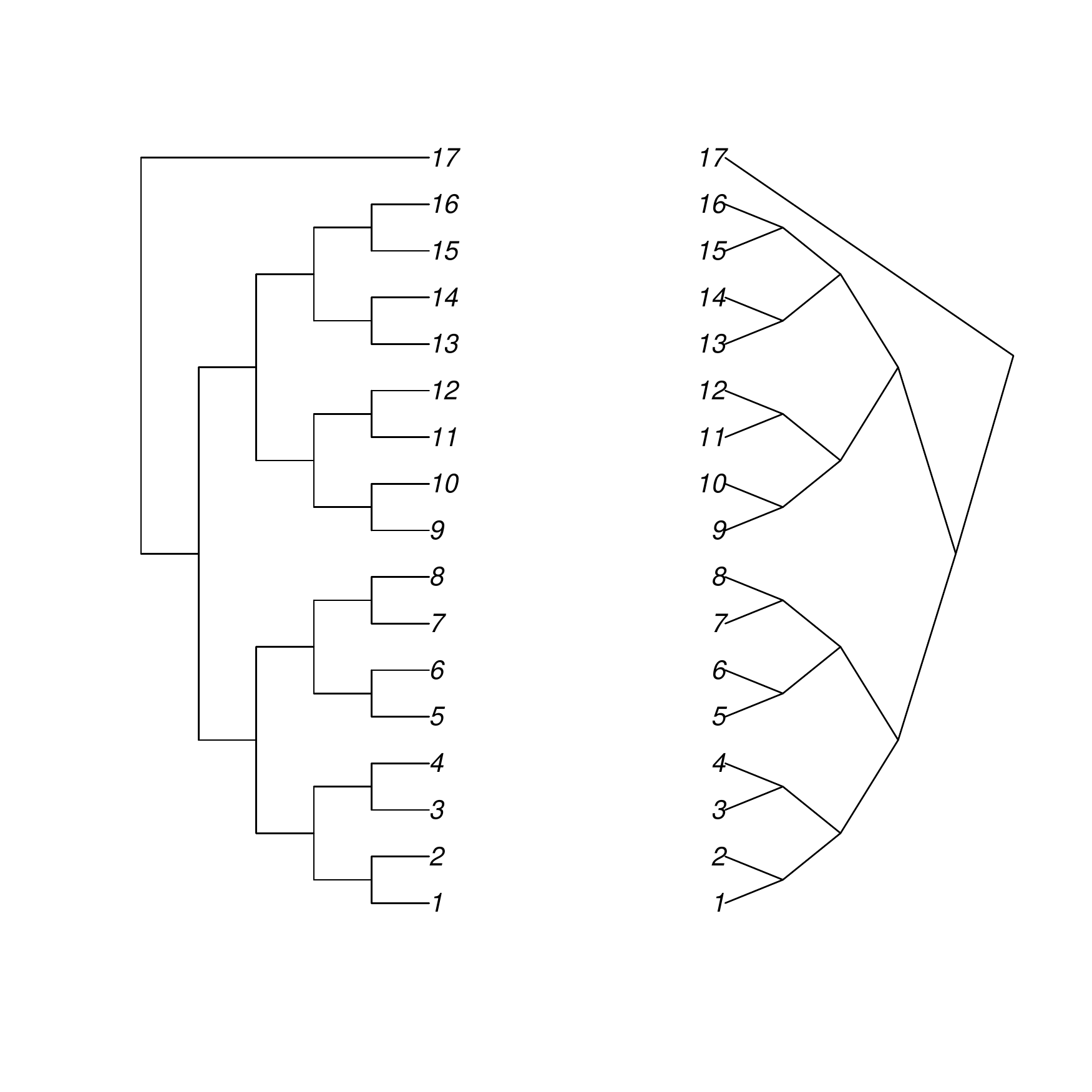} 
   \vskip-1cm
   \caption{Two common representations of trees. The dendrogram
     representation of a balanced tree with 17 leaves is on the left, the
     cladogram representation of the same tree is on the right.}
   \label{fig:tree16}
\end{figure}

On the right, the cladogram is often considered a useful representation of
an evolutionary process, where the root represents the common ancestor to
the 17 leaves or taxa at the end of the tree.  The simplest evolutionary
process is that denoted by the Cavender-Farris-Neyman \citep{felsenstein} process where each
position or character is binary.  Suppose we consider just one character;
we generate the character at the root at random, say from a fair
coin. This character will then `evolve' (that is, be pushed) through the
tree from the root to the leaves, with some probability of mutating as it
passes through edges. The mutations have a higher probability of occurring
over longer branches. The simplest model for mutation is to suppose a
molecular clock, and thus that the rate is the same throughout the tree
and always proportional to the edge lengths.  If the mutation rate is very
low, we might end up with all the characters at the 17 leaves equal to the
root, while if it is very high, the characters at the leaves will have
very little resemblance to those at the root. The leaves usually represent
the contemporary taxa and thus we can guess what was at the root by what
occurs at the leaves as long as the mutation rate is neither too low nor
too high.
\\
\begin{figure}[h] 
   \centering
\includegraphics[width=3.4in]{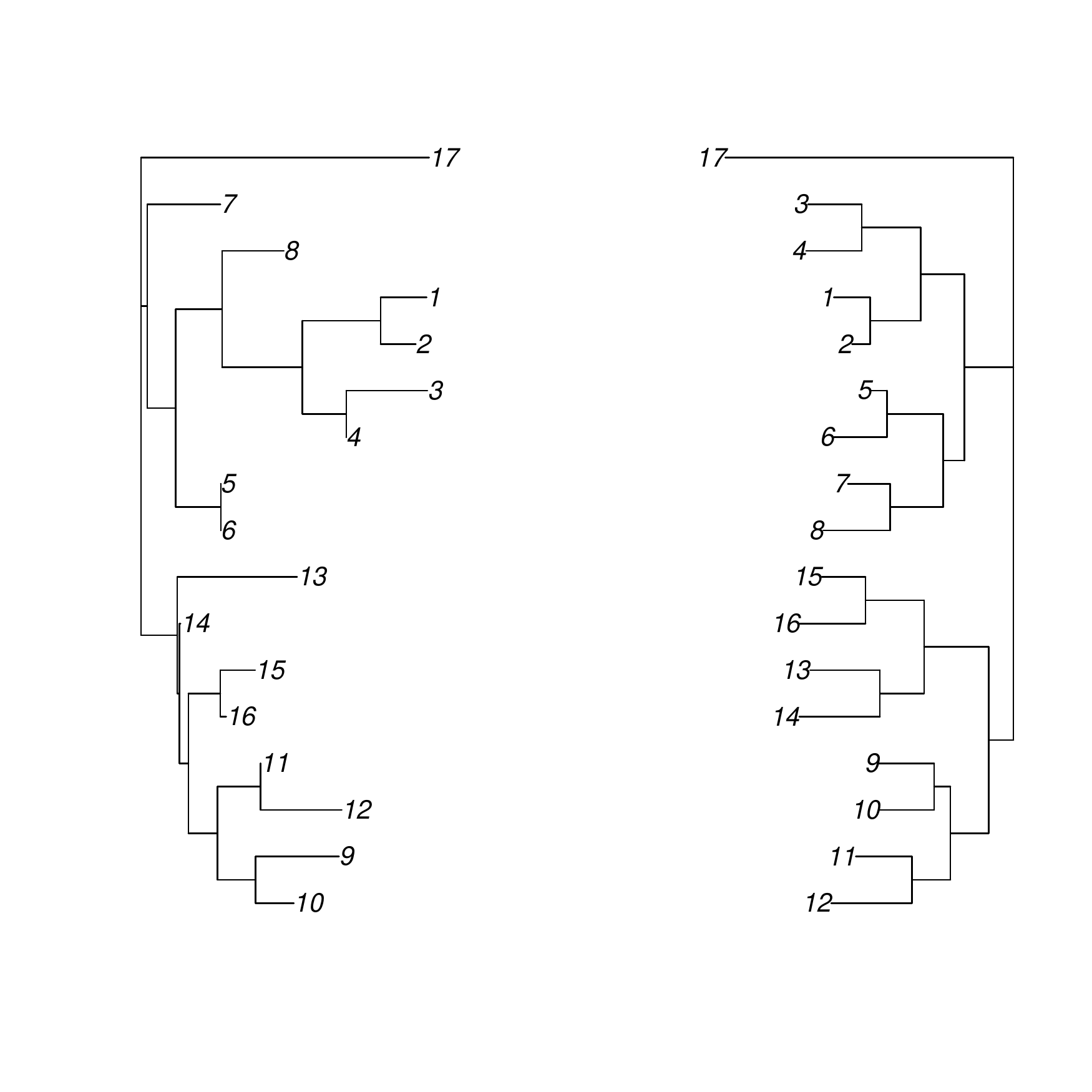}
\vskip-1cm
\caption{The tree on the left was estimated with a matrix which had 100
  columns (characters), of which only 24 different patterns are
  represented in the columns. The tree on the right was estimated with a
  matrix of 400 columns, with 68 different patterns represented. We can
  see that the tree on the right has the correct branching order, as
  compared to the estimate on the left.  }
   \label{fig:compare16}
\end{figure}

Another factor that effects the tree estimation quality is tree shape. As
an example, here are two trees that were used to generate data. We call
the left one the comb tree and the right one the balanced tree.
  \begin{figure}[h!] 
   \centering
\includegraphics[width=3.6in]{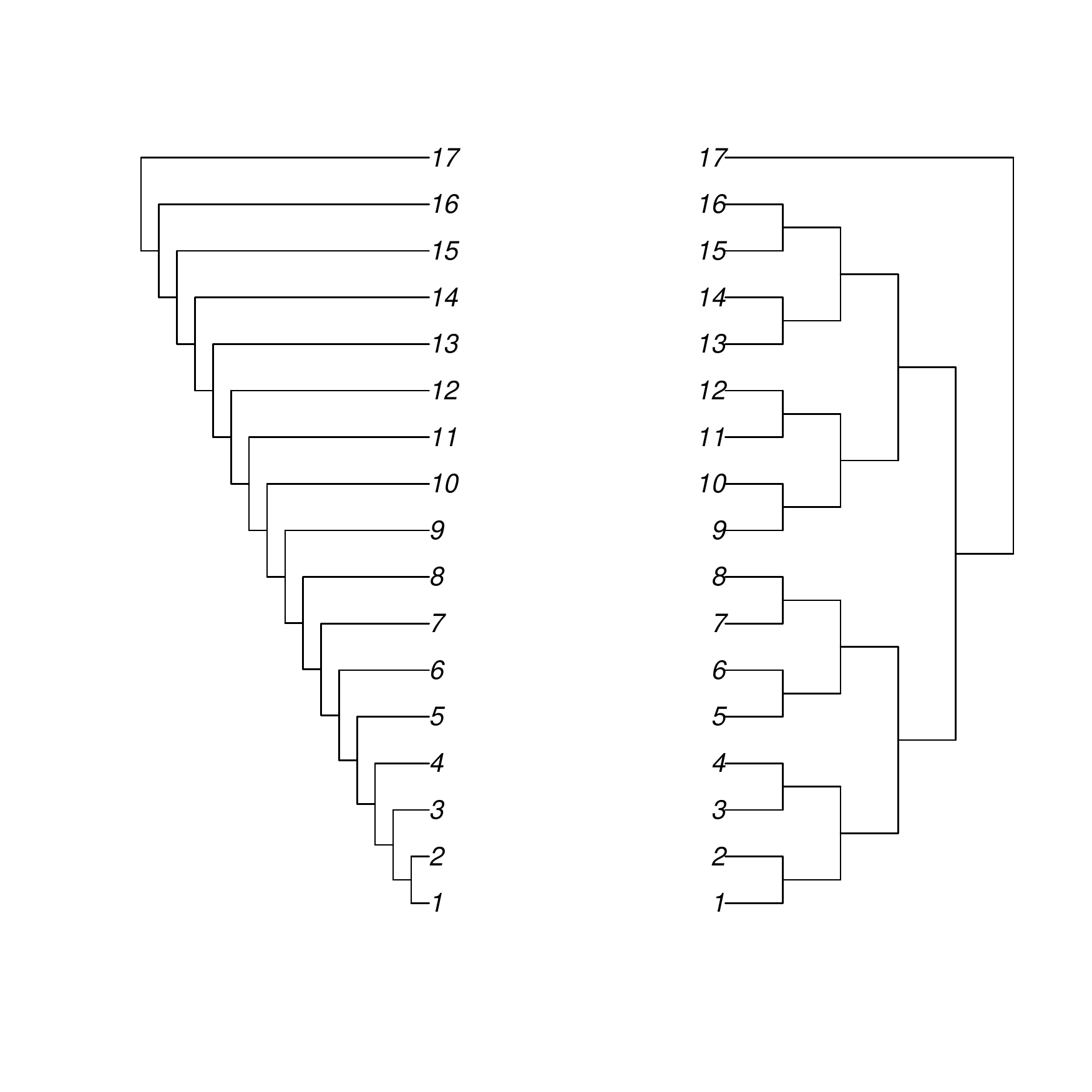}
\vskip-2cm
\caption{Two theoretical trees were used to simulate sequences of length
  400. The tree on the left is the comb tree, and the tree on the right is
  often called the balanced tree. }
   \label{fig:compare16theory}
\end{figure}
 \begin{figure}[h!]
   \centering
\includegraphics[width=4in]{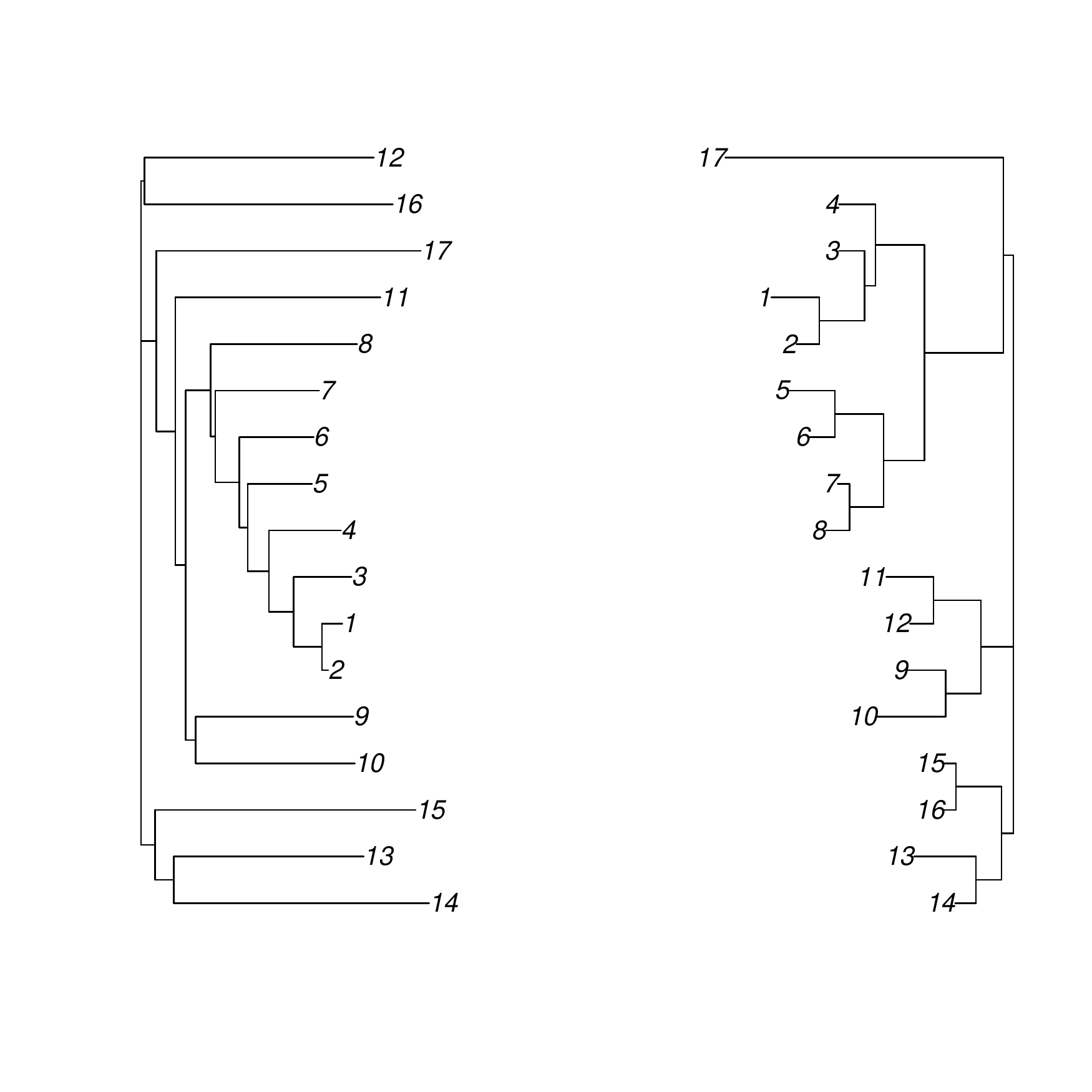}
\vskip-2cm
\caption{The two theoretical trees from Figure \ref{fig:compare16theory}
  were used to simulate sequences of length 400. From this example it
  seems the tree with the larger inner branch lengths was easier to
  estimate. We will make this precise in Section 3. }
   \label{fig:compare16data}
\end{figure}

Tree estimation methods can be grouped into categories of parametric,
non-parametric, and intermediate methods. One popular parametric
estimation method is maximum likelihood (MLE) (a classical textbook
presentation of this can be found in \cite{felsenstein}). Although this is
known to be NP-complete, remarkable computational advances have been made
in recent years with regards to this reconstruction problem and
particularly efficient implementations exist, such as RAxML   \citep{RaxML}
or PhyML   \citep{phyml}. Another parametric approach is the Bayesian
Maximum A Posteriori (MAP) estimate, implemented in \texttt{Mr Bayes}
  \citep{MrBayes} or \texttt{Beast} \citep{Beast} following Yang and
Rannala's prescription for Bayesian inference and tree estimation
  \citep{BayesTrees}.
 
The most standard nonparametric estimate for a tree is the Maximum
Parsimony tree. The reconstruction algorithm is designed to take the original
data as the leaves and add extra ancestral points  that minimize the
number of mutations needed to explain the data. From a computational
viewpoint, this is equivalent to the Steiner tree problem and is known to
NP-complete   \citep{NP}.

Intermediaries between strictly parametric methods based on a finite
number of mutational rate-parameters and an evolutionary tree and the nonparametric
approaches are the distance based methods. These use the parametric
Markovian models to find the distances between species and then lean on
various heuristic criteria to build a binary tree from these distances.

Distance based methods forfeit the use of the individual columns in favor
of a simple distance between the aligned sequences, although recent work
by \cite{Roch} shows there might not be such an important loss in
information. The distance can use any of the standard Markovian
evolutionary models such as the Jukes Cantor one parameter model or Kimura
two parameter model   \citep{felsenstein}, or a simple Hamming distance
between sequences.  Once the distances have been computed the tree
building procedure can follow one of many possible heuristics.  Neighbor
Joining is an agglomerative (bottom up) method which is computationally
inexpensive and is thus often used as a starting point for other tree
estimation procedures.  UPGMA or average linkage method is another such
method that updates the distance between two clusters by taking the
average distance between all pairs of points from the two groups (the
orignal idea is explained in \cite{MichenerSokal}, a standard treatment
can be found in \cite{felsenstein}).

\subsection{Hierarchical clustering trees}
Building hierarchical clustering trees is very similar to the use of
distances to build phylogenetic trees, with the difference that the choice
of distances or even of simpler dissimilarities between the leaves is no
longer driven by an evolutionary model but dissimilarities either in gene
expression or in occurrence of words or other relevant
features  \citep{Hartigan:1967}.  The resulting hierarchical clustering has
the advantage over simple clustering methods such as $k$-means that one
can look at the output in order to make an informed decision as to the
relevant number of clusters for a particular data set.

\subsection{Methods for Evaluating Trees}
In the case of hierarchical clustering trees, \cite{fowlkesm} provide a
first approach to comparing two hierarchical clusterings by creating a
weighted matching score from the matrix of matchings.  However, recently
more global distributional assessments of clusterings of trees have been
possible thanks to advances in computation:
\begin{description}
\item[Bootstrap support for Phylogenies] As in the standard bootstrap
  technique \cite{Efron}, with observations being the columns of the
  aligned sequences, the sampling distribution of the estimated tree is
  estimated by resampling with replacement among the characters or columns
  of the data.  This provides a large set of plausible clusterings. These
  were used for instance by \cite{felsen} to build a confidence statement
  relevant to each split.  An adjustment was proposed in
  \cite{efronh}. This is based on finding a path between trees that are
  each side of the boundary separating two tree topologies; we show in
  Section 4 how this can be implemented using our implementation of the
  geometric distance.
\item [Parametric Bootstrapping for Microarray Clusters]
  \cite{KerrChurchill} have proposed a way of validating hierarchical
  clustering as it is used in microarray analysis.  Their model is a
  parametric ANOVA model for microarrays which includes gene, dye and
  array effects. Once these effects have been estimated on the data,
  simulated data incorporate realistic noise distributions can be
  generated through a parametric bootstrap type procedure.  From the
  simulated data many hierarchical clusters are generated and then
  compared.  The authors use this to evaluate the stability of a gene,
  using percent of bootstrap clusterings in which it matches to the same
  cluster in the same way \cite{felsen} provides the estimate of the
  binomial proportion of trees with a given clade.  We can repeat their
  generation process but again combine the trees differently than by
  estimating a presence-absence estimate for each clade.  We show in
  section 3 how a more multivariate approach can provide richer
  visualizations of the stability of hierarchical clustering trees.
\item[ Bayesian posterior distributions for phylogenetic trees]
  \cite{BayesTrees} develop the Bayesian framework for estimating
  phylogenetic trees using a Bayesian posterior probability distribution
  to assess stability.  The usual models put prior distributions on the
  rates and a uniform distribution on the original tree and then proceed
  through the use of MCMC to generate instances of the posterior
  distribution.  Since implementations such as \cite{MrBayes} provides a
  sample of trees from the posterior distribution, these can be used for
  the same purpose as the bootstrap resample of trees.  Following
  procedures exposed in section 3, we can combine these picks from the
  posterior distribution using the distances to give an estimate of a
  median posterior tree and to give multivariate representations such as
  hierarchical clusters and MDS plots of the posterior distribution.
\item[ Bayesian methods in hierarchical clustering] \cite{Heller} provide
  a Bayesian nonparametric method for generating posterior distributions
  of hierarchical clustering trees.  Visualizing such posterior
  distributions can be tackled with the same tools as those used for
  Bayesian phylogenetics.
\end{description}


\section{The Polynomial Time Geodesic Path Algorithm}

\subsection{Path Spaces, Geodesics, and Uniqueness}
The distance algorithm implemented computes the geodesic distance metric
proposed by \citet*{Bhv}.  This arises naturally from their formulation of
tree space as a space made up of Euclidean orthants. A path between two
trees consists of line segments through a sequence of orthants. This
sequence of orthants is the \emph{path space}. A path is a \emph{geodesic}
when it has the smallest length of all paths between two points.

As \citet*{Bhv}  showed, tree space is a negatively curved CAT(0) space.
 As a consequence, there is a unique
geodesic between any two trees  \citep{Gromov}. We then can find the
distance between two trees by finding the geodesic path.

\subsection{The Algorithm Intuitively}
 \cite{owenprovan} have proposed an
iterative method to constrict a path until it is the geodesic between
its two endpoints. Since all orthants connect at the origin, any two
trees $T$ and $T'$ can be connected by a two-segment path,
consisting of one segment from $T$ to the origin, and another from
the origin to $T'$. This path is in general not a geodesic, but is an
easily definable path that must always be valid, making it a useful
starting point. This path is called the \emph{cone path}.

From this start, the algorithm iteratively splits a transition between
orthants into two transitions by introducing a new itermediate orthant
into the path space until a condition is met such that we know the
path is the geodesic. We then compute the length of the path to get
the geodesic distance between the two trees.

\subsection{Notation and Setup}
Let $T$ and $T'$ be two rooted semi-labeled weighted binary trees with $n$
labeled tips, with labels $X = 1..n$, and $2n-2$ edges $\calE$ and
$\calE'$. Formally, we hang the trees by a labeled root $Z$, though
because of a representational trick it is not necessary to include this in
the computations. Every edge $e \in \calE$ defines a partition of labels
$X_e | \overline{X}_e$. As a convention we will consider $X_e$ as the set
of labels `below' $e$ (that is, the set not including $Z$), and
$\overline{X}_e$ as the set of edges `above' or `rootward' of $e$, or the
set containing $Z$.

Two edges $e$ and $f$ are called \emph{compatible} if one of $X_e \cap
X_f$, $X_e \cap \overline{X}_f$, $\overline{X}_e \cap X_f$, or
$\overline{X}_e \cap \overline{X}_f$ is empty. Intuitively, edges are
incompatible if they could not both be edges in the same tree. The
partition formed by an edge uniquely identifies it amongst all edges
in $n$-trees, so we represent each edge by the partition it forms, and
call two edges the same if they form the same partition of $X$. Two
sets of edges $A$,$B$ are compatible if for every $e \in A$, $e$ is
compatible with every $f \in B$.

We represent a path between two trees by $(\mathcal{A}, \mathcal{B})$ with
$\mathcal{A} = (A_1, ..., A_k)$ and $\mathcal{B} = (B_1, ..., B_k)$, where
$A_i \subseteq \calE$ represents the edges dropped from $T$, and $B_i$
represents the edges added from $T'$ at the $i^{\text{th}}$ transition
between orthants. The norm $||A||$ of a set of edges is $\sqrt{\sum_{e \in
    A}|e|^2}$ where $|e|$ is the weight of $e$.

\subsection{Conditions for a Path to Be a Geodesic}
\cite{owenprovan} give
two properties that all valid
paths must fulfill, as well as a third property that fully
characterizes a geodesic path. The properties are presented here
without development. 

\begin{theorem}{A sequence of sets of edges $(\mathcal{A}, \mathcal{B})$
  represents a valid path if and only if (1) for each $i > j$, $A_i$
  and $B_j$ are compatible, and (2) $\frac{||A_1||}{||B_1||} \le
  \frac{||A_2||}{||B_2||} \le ... \le \frac{||A_k||}{||B_k||}$.}
\end{theorem}

\begin{theorem}{(Property 3) A valid path is the geodesic path if and only
  if, for each $(A_i, B_i)$ in the path, there is no partition $C_1
  \cup C_2$ of $A_i$ and partition $D_1 \cup D_2$ of $B_i$ such that
  $C_2$ is compatible with $D_1$ and $\frac{||C_1||}{||D_1||} \le
  \frac{||C_2||}{||D_2||}$.}
\end{theorem}

\cite{owenprovan} proved that properties (1) and (2) are always
satisfied throughout the algorithm, and present a polynomial time way
to both check (3) and split $(A_i,B_i)$ into $(A_i,B_i),(A_{i+1},
B_{i+1})$.

\subsection{Representing Trees, Preprocessing, and Optimizations}
Trees are uniquely represented by the set of partitions formed by all
edges in the tree. We represent an edge by an $n+1$-length vector of
logical values, a true value meaning the leaf in that position is
`downward' of the edge, with the $n+1$ position representing the root node
$Z$. This representation allows efficient computation of edge
compatibility (i.e. in O(n) time, since one iteration through the vectors
can check all intersections).

Before the  algorithm is implemented, several
preprocessing steps are performed. There are three preprocessing
steps: (1) edges are classified as shared between the two trees or
unique to its own tree, (2) the trees are divided up into
independently solvable subproblems, and (3) edge incompatibility
information is computed and cached. 

The first step, classification of edges as shared or unique, serves
two purposes. Edges shared by the two trees will not get dropped or
added by any transition through an orthant, and so we can think of all
the shared edges as having Euclidean distance within one
orthant. Since they can be added in as such at any time, there is no
purpose in using them in the high-cost distance computation, and
indeed the algorithm presented by Owen and Provan requires trees to be
disjoint.

The second purpose this classification serves is to aid in the second
preprocessing step, division into subproblems. The division works from
the observation that for every shared edge in $T$,$T'$, we can treat
the distance between the subtrees from this edge as simply part of the
distance between the two edges.  This lets us compute the distance
between these two subtrees in an identical way to the larger tree, and
integrate this distance as if it were a shared edge distance. 

This division is accomplished by classifying every unique edge in both
trees under the tightest shared edge, i.e. working upwards until we
reach a shared edge. Computationally this takes the form of, for every
edge $e$, computing the difference in number of downward leaves
between every shared edge and $e$, and taking the shared edge with the
minimum positive difference. This approach is used since no
representation of the tree as an actual binary tree is stored. The
computation is still reasonably efficient due to the vector
representation of the partitions and that the count of downward edges
can be cached.

From the second step, we get a series of bins, each containing a pair of
disjoint subtrees from a shared edge root. For each bin, we compute all
pairwise edge compatibilities between edges on the two trees, caching
their result in a vector of edge pairs. While these implementation details
do not affect the theoretical efficiency of the algorithm, they make the
computation reasonable in practice for large datasets.

\subsection{The Algorithm}
The algorithm itself reduces to checking property (3). Following
\cite{Staple}, the notion of incompatibility between edges is coded into a
bipartite graph. Owen and Provan show that property (3) can be checked by
forming a graph $G(A_i,B_i)$ and computing the minimum weight vertex cover
for this graph. The graph $G(A_i,B_i)$ is formed by adding graph edges
between incompatible edges of $A_i$ and $B_i$, and weighting each vertex
$v \in A_i$ as $\frac{||v||^2}{||A_i||^2}$ and $w \in B_i$ as
$\frac{||w||^2}{||B_i||^2}$. Because this graph is bipartite, the minimum
weight vertex cover problem can be solved using a max-flow algorithm with
the following conversion: source node $s$ and sink node $t$ are added to
the graph, edges are added between $s$ and every $v \in A_i$ with edge
weight equal to the vertex weight of $v$, the edges given by
incompatibilities are given infinite weight, and edges are added between
every $w \in B_i$ and $t$ with weight equal to the vertex weight of
$w$. The Edmonds-Karp max-flow algorithm was used to compute the max flow,
giving a runtime complexity of $O(|V|\cdot |E|^2)$, where $|V| = |A_i| +
|B_i| + 2$, and $|E| \le |A_i| + |B_i| + |A_i|\cdot |B_i|$. For a subtree
with $k$ unique edges and $n$ leaves, $|A_i|, |B_i| \le k \le n-2$, giving
a worst-case complexity of $O(n^3)$ for checking property (3). A breadth
first search of the residual flow graph, after dropping 0-weight edges,
gives the subsets $C_1 \subset A_i$ and $D_1 \subset B_i$ ($C_1 \cup
\overline{D_1}$ happens to be the minimum weight vertex cover, but at this
point we don't need the actual cover itself).

For each bin, we run the algorithm.  Initialize $(\mathcal{A},
\mathcal{B})$ to $A_0 = \calE$ and $B_0 = \calE'$, that is, all edges
dropped and added in one orthant transition (this is effectively the cone
path of the subtree, which we know to be a valid path, satisfying
properties (1) and (2)). Iteratively, we run the following procedure on
all pairs $(A_i, B_i)$ of $(\mathcal{A}, \mathcal{B})$: (1) compute the
max-flow $f$ of $G(A_i, B_i)$ (2) if $f < 1$, find all accessable edges
$C_1 \subset A_i$ and $D_1 \subset B_i$ and replace $(A_i,B_i)$ with
$(C_1,D_1),(\overline{C_1},\overline{D_1})$ in $(\mathcal{A},
\mathcal{B})$. When, for each pair $(A_i,B_i)$, the max-flow of $G(A_i,
B_i)$ is $\ge 1$, $(\mathcal{A}, \mathcal{B})$ represents the geodesic
path for the subtree, and the algorithm is done. Since at each step,
$|A_i|$ and $|B_i|$ are greater than or equal to $1$ and we cannot add or
remove more edges than the trees have, we have $O(n)$ iterations of the
above steps, giving a total running time of $O(n^4)$.

As an alternative to the Edmonds-Karp algorithm, a linear programming
solution, simpler than a standard linear program for bipartite max flow,
can be used. In our experiments, it appears to be more computationally
efficient for small $n$, however, it has asymtotically worse
performance. For very small $A_i$ and $B_i$ sets, a brute force solution
may be more computationally efficient, since all choices of a minimum
weight vertex cover may be able to be enumerated more efficiently than
setting up the machinery for a max-flow algorithm.

\subsection{Final Calculation}
Once all geodesic paths between subtrees are found, we compute the
final distance. Denote the paths between subtrees by $\mathcal{P}$,
and the set of shared edges $\mathcal{S}$. Then the final distance
between the two trees is given by
\begin{equation*}
  d(T,T') = \sqrt{ \sum_{(A_i, B_i) \in \mathcal{P}}
    \left( ||A_i|| + ||B_i|| \right)^2
    + \sum_{s \in \mathcal{S}}\left(|s_T| - |s_{T'}|\right)^2}
\end{equation*}

\subsection{Implementation}
The algorithm has been implemented in the R package {\tt distory}
  \citep{distory}, containing many of the examples from this paper. The
package requires the {\tt ape}   \citep{cran-ape} package for analyzing
phylogenetic trees and can be beneficially supplemented by the {\tt
  phangorn}   \citep{phangorn} package.

The current implementation in {\tt distory} can compute all
pairwise distances between 200 bootstrap replicates of a 146-tip tree in
approximately 2 minutes seconds on a Core 2 Duo 1.6ghz processor. 

\section{Choosing a geometry for embedding trees}
\subsection{Non positively curved spaces}
\citet*{Bhv} show that the distance as computed above endows the space of
trees with a negative curvature.  See the excellent book length treatment
of non-positively curved metric spaces in \cite{BrHa}.

\begin{figure}[htbp] 
\hskip-0.5cm
   \includegraphics[width=7in]{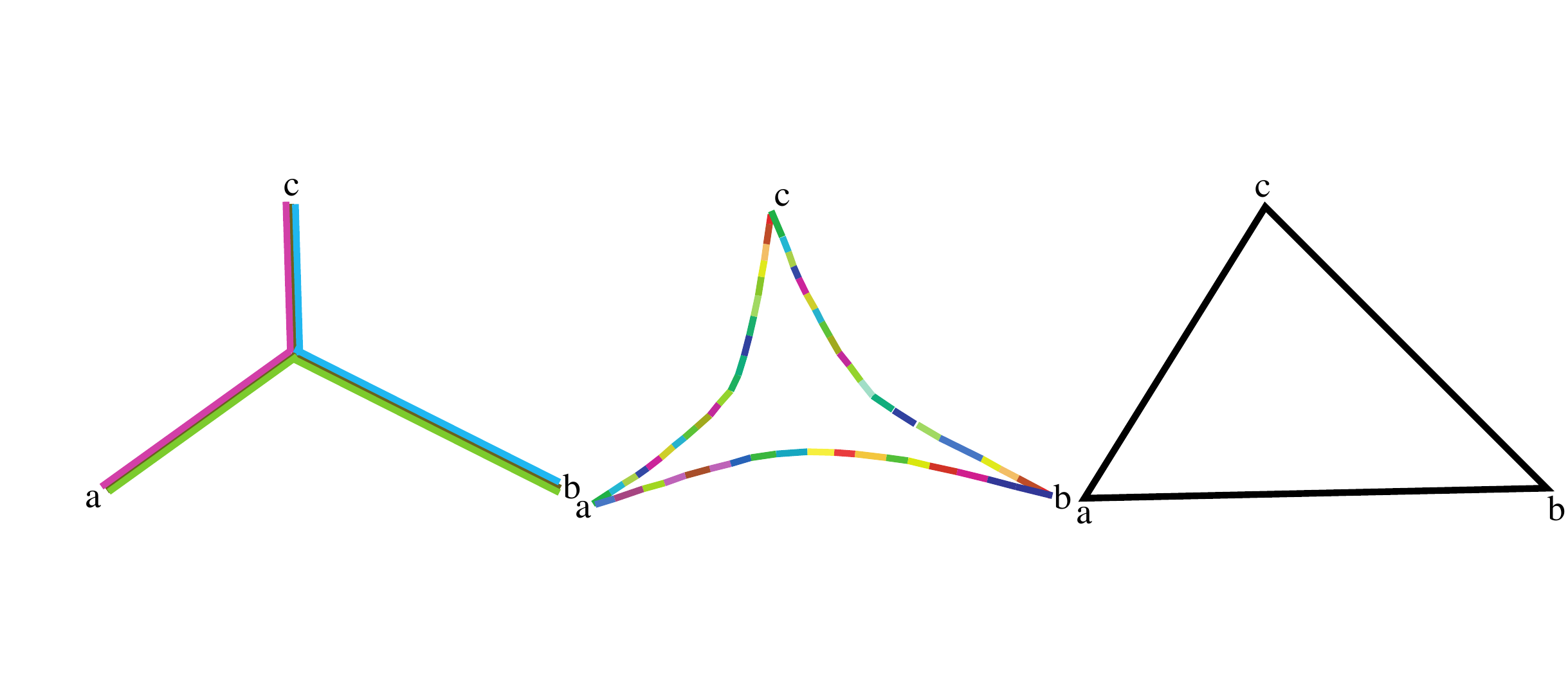} 
   \vskip-2cm
   \caption{Three triangles illustrating non-positively curved spaces. The
     center one represents three points (trees) in treespace, a, b and c,
     with the geodesics running between them (notice the paths are made of
     sequences of segments that sit in the Euclidean cubes of the cube
     complex, but we can see an overall negative curvature, i.e. triangles
     are thin compared to the Euclidean comparison triangle on the right).
     The left triangle depicts the situation in which the space is so
     negatively curved as to be a tree. }
   \label{fig:threetriangles}
\end{figure}

This is illustrated by Figure~\ref{fig:threetriangles}, which shows three
geodesic triangles in three different
types of space.  On the left, a $\delta$-hyperbolic metric space with $\delta$=0, is actually a tree.\\
\specialbox{$\delta$-hyperbolic metric space }{
\begin{enumerate}
\item Consider a geodesic triangle: 3 vertices connected by geodesic
  paths.  It is $\delta$-thin if any point on any of the edges of the
  triangle is within distance $\delta$ from one of the other two sides.
\item
A
 $\delta$-hyperbolic space is a geodesic metric space in which every geodesic triangle is 
  $\delta$-thin.
\end{enumerate}
} As we can see in Figure~\ref{fig:threetriangles}, the `triangle in a
tree' on the left is represented by the three colored edges (made of two
segments each) and has the property that each point from an edge, for
instance a point on the pink edge $ac$ is within $\delta=0$ from the
closest other edge, either the blue or the green are at distance $0$.  The
middle triangle has a $\delta=0.25$ (if we think the long side $d(a,b)$ is
1), and the right triangle has a $\delta=0.5$.  Euclidean space actually
does not have a bounded $\delta$, ($\delta=\infty$), as triangles can be
chosen to be arbitrarily large.

The question raised by Figure~\ref{fig:threetriangles} and that we will
try to address is whether we can make the best approximate representation
of many trees given their BHV distances by embedding the points in a
Euclidean space using a modified MDS or whether it is better to place the
trees in a tree, as we do in the last section of the paper.  The question
of choice between spatial and treelike representations is an old one and
was clearly posed by \cite*{Pruzansky} almost 30 years ago in the context
of dissimilarities measured on psychological preferences.  We introduce a
more quantitative notion by computing the $\delta$-hyperbolicity of a set
of distances between a finite set of points.  The question of the quality
of a treelike approximation is considered in section
\ref{sec:treeoftrees}. In the next section we will show how to measure the
quality of a Euclidean approximation.
\subsection{Multidimensional Scaling and its application to tree comparisons}
Psychometricians, ecologists and statisticians have long favored
a method known as multidimensional scaling (MDS) to
approximate general dissimilarities with Euclidean distances.

MDS is a statistical method developed around Schoenberg's theorem that a symmetric matrix of positive entries with zeros on the diagonal is a distance matrix between $n$ points if and only if the matrix
$$-\frac{1}{2}HD^2H \mbox{ is positive semi-definite }$$
We will not provide the details of the algorithm, referring the reader to
\cite*[p.407]{Mardia} but offer the following summary:
Given an $n \times n$ matrix of interpoint distances, one can solve for  points 
in Euclidean space approximating  these distances by:
\begin{enumerate}
\item Double centering the interpoint distance squared matrix: $S = -\frac{1}{2}H D^2 H$.
\item Diagonalizing $S$: $S = U \Lambda U^T$.
\item Extracting $\tilde{X}$: $\tilde{X} = U \Lambda^{1/2}$.
\end{enumerate}
We use the standard implementation provided in the \texttt{stats} package of \texttt{R} \cite{R} by the function
\texttt{cmdscale}. 
We can estimate the quality of the approximation of the distances $d$ using the Euclidean
approximation $ $ by computing an index such as
$$
 \sum_{i<j}(d_{ij}-\delta_{ij})^2
.$$
This index will be zero if the data come from a $k$ dimensional Euclidean space and
we retain as $k$ dimensions in our multidimensional scaling.

As we will see later, since we know our tree points lie on a curved space where local
distances  are nearly Euclidean but points further apart are not,
we will  sometimes be better off making modifications to the distances
before applying MDS.
\subsection{MDS of Bootstrapped Trees}
One approach to inference for hierarchical clustering and phylogenetic trees is to simply apply a nonparametric
resampling bootstrap to the data and re-estimate the trees. This gives an idea of the overall
variability of the data under the assumption that the unknown distribution of the distances
$d(\tau,\hat{\tau})$ can be well approximated by that of
$d(\hat{\tau},\hat{\tau}^*)$ , where $\hat{\tau}^*$ denotes the bootstrapped estimates of the tree.

\begin{figure}[h!] 
   \centering
   \includegraphics[width=5in,height=4in]{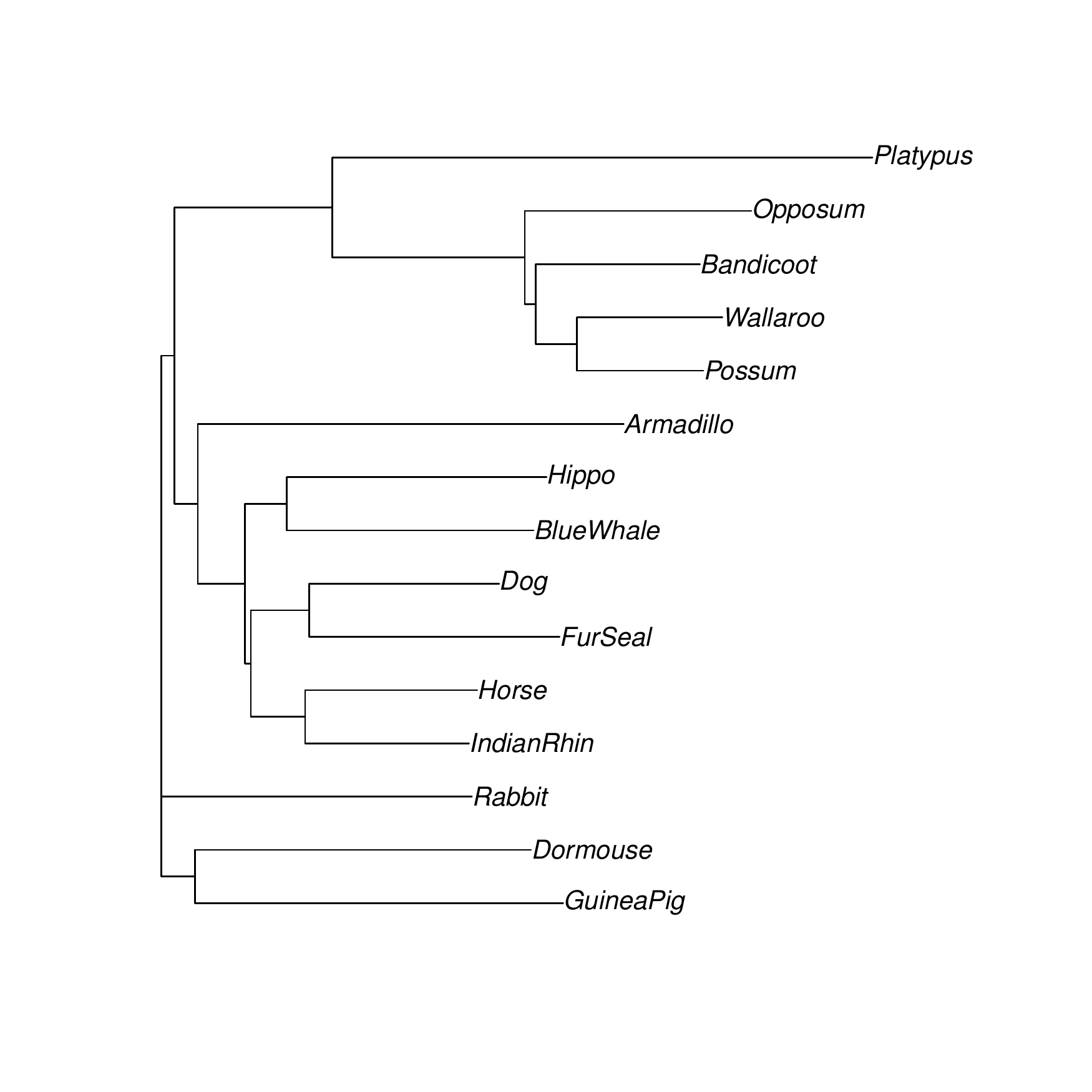} 
   \vskip-2cm
   \caption{Tree Estimate from  aligned DNA sequences of length 3179.}
   \label{fig:laurtree}
\end{figure}

Here we apply this idea in conjunction with a MDS plot, using a bootstrap
of the the Laurasiatherian DNA data from the package \texttt{phangorn}
  \citep{phangorn}. The original estimate is shown in
Figure~\ref{fig:laurtree}.

\begin{figure}[ht] 
   \centering
   \includegraphics[width=6in]{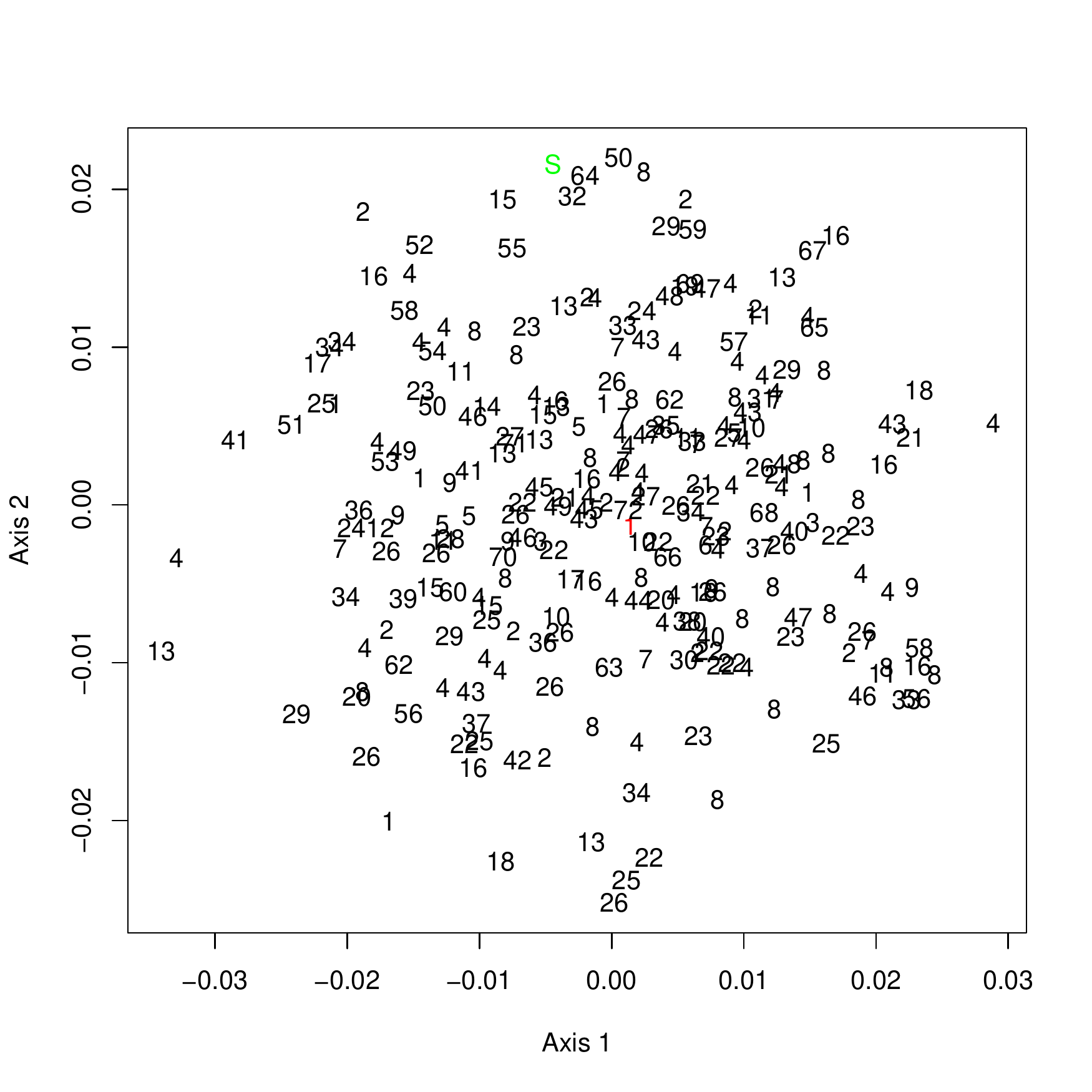} 
   \caption{First MDS plane representing 250 bootstraps. The tree
     topologies were numbered from 1 (the original tree) to 72. We have
     plotted the orginal bootstrap estimate in red as 1, and the star tree
     with the pendant edge lengths matching those of the original tree
     represented by the letter S in green.}
   \label{fig:bootmds}
\end{figure}

\begin{center}
\begin{table}[ht]
\begin{tabular}{|rr|}
  \hline
Tree type & count \\ 
  \hline
1 &   6 \\ 
  2 &  12 \\ 
  3 &   2 \\ 
  4 &  37 \\ 
  5 &   4 \\ 
  6 &   1 \\ 
  7 &   9 \\ 
  8 &  21 \\ 
  9 &   5 \\ 
  10 &   3 \\ 
  11 &   5 \\ 
  12 &   1 \\ 
  13 &   7 \\ 
  14 &   1 \\ 
  15 &   4 \\ 
  16 &   8 \\ 
      17 &   2 \\ 
  18 &   2 \\ 
  \hline
  \end{tabular}
  \begin{tabular}{|rr|}
  \hline
Tree type & count \\ 
  \hline
  19 &   1 \\ 
  20 &   3 \\ 
   21 &   4 \\ 
  22 &  10 \\ 
  23 &   6 \\ 
  24 &   2 \\ 
  25 &   5 \\ 
  26 &  14 \\ 
  27 &   3 \\ 
  28 &   1 \\ 
  29 &   4 \\ 
  30 &   1 \\ 
  31 &   1 \\ 
  32 &   1 \\ 
   33 &   3 \\ 
  34 &   5 \\ 
  35 &   1 \\ 
  36 &   2 \\ 
   \hline
\end{tabular}
  \begin{tabular}{|rr|}
  \hline
Tree type & count \\ 
\hline
  37 &   2 \\ 
  38 &   1 \\ 
  39 &   1 \\ 
  40 &   2 \\ 
  41 &   2 \\ 
  42 &   1 \\ 
  43 &   5 \\ 
  44 &   1 \\ 
  45 &   2 \\ 
    46 &   3 \\ 
  47 &   3 \\ 
  48 &   2 \\ 
   49 &   2 \\ 
  50 &   2 \\ 
  51 &   1 \\ 
  52 &   1 \\ 
  53 &   1 \\ 
  54 &   1 \\ 
     \hline
\end{tabular}
  \begin{tabular}{|rr|}
  \hline
Tree type & count \\ 
\hline
   55 &   1 \\ 
  56 &   2 \\ 
  57 &   1 \\ 
  58 &   2 \\ 
  59 &   1 \\ 
  60 &   1 \\ 
  61 &   1 \\ 
  62 &   2 \\ 
  63 &   1 \\ 
  64 &   1 \\ 
  65 &   1 \\ 
  66 &   1 \\ 
  67 &   1 \\ 
  68 &   1 \\ 
  69 &   1 \\ 
  70 &   1 \\ 
  71 &   1 \\ 
  72 &   1 \\ 
   \hline
\end{tabular}
\vskip0.5cm
\caption{Binned Bootstrapped Trees. Of the 250 bootstrapped trees, there is a majority of type 4.}
   \label{tab:binLaura}
\end{table}
\end{center}
Table \ref{tab:binLaura} shows
that the 250 trees are of 72 different types or branching orders.
An MDS plot of the first two principal coordinates using the BHV distance
is presented in Figure~\ref{fig:bootmds}.

If we compute a simple Shannon type diversity
index, we get the impression that the trees are very diverse
(SW=$-\sum
p_i\log(p_i)-\frac{S-1}{2*N}$=\texttt{-sum(freq*log(freq))-(71/502)=3.5}).

In fact the trees as represented in Figure~\ref{fig:bootmds} are quite
grouped.

It is an interesting question how to study the variability of trees,
whether using a diversity index or an `inertia' type approach using such
sums of squares of distances.  The original estimate projects at the
center of the scatterplot, leading us to believe that the estimate is
unbiased.

As an additional element we have projected the star tree ``S" (chosen with
the lengths of the pendant edges closest to the original tree) to see
whether it is in a small neighborhood, or credibility region of the
bootstrapped trees. This is analogous to seeing if $0$ is in a confidence
interval of differences between two random variables.  If the star tree
seems to be in a confidence region with a high probability coverage then
we may conclude that the data are not really treelike. In
Figure~\ref{fig:bootmds}, $S$ appears to be on the outer convex hull of
the projected points; we can conclude that the probability that the star
tree belongs to the confidence region is low. See \cite{ihp} for details
on the idea of using convex hulls to make confidence statements of this
type.

Looking at Figure~\ref{fig:bootmds}
we can see
that trees of the same topology are not necessarily closer to the original
tree if we use the {\bf BHV} with no modifications. In some cases we may want to give an extra weight
to crossing orthants. We give  examples of such modifications of the distance in the \cite{distory}
vignette.
\subsection{Empirical Evidence of Mixing on the  Bethe Lattice}
\cite{ErdosSteel} have shown that the tree shape that requires the longest
sequence length for inferring the root as accurately as possible is the
balanced tree.  \cite{Mossel:2004} recognized this tree shape as the Bethe
Lattice, known in statistical mechanics, and used this fact to give bounds
on the sequence lengths necessary to rebuild the tree accurately with a
given probability.  For this shape, \cite{Mossel:2004} showed that if
mutation rates are high it is impossible to reconstruct ancestral data at
the root and the topology of large phylogenetic trees from a number of
characters smaller than a low-degree polynomial in the number of leaves.
\\
\begin{figure}[ht]
\vskip-0.67cm
\hskip-2.5cm
\includegraphics[width=7.2in,height=4.3in]{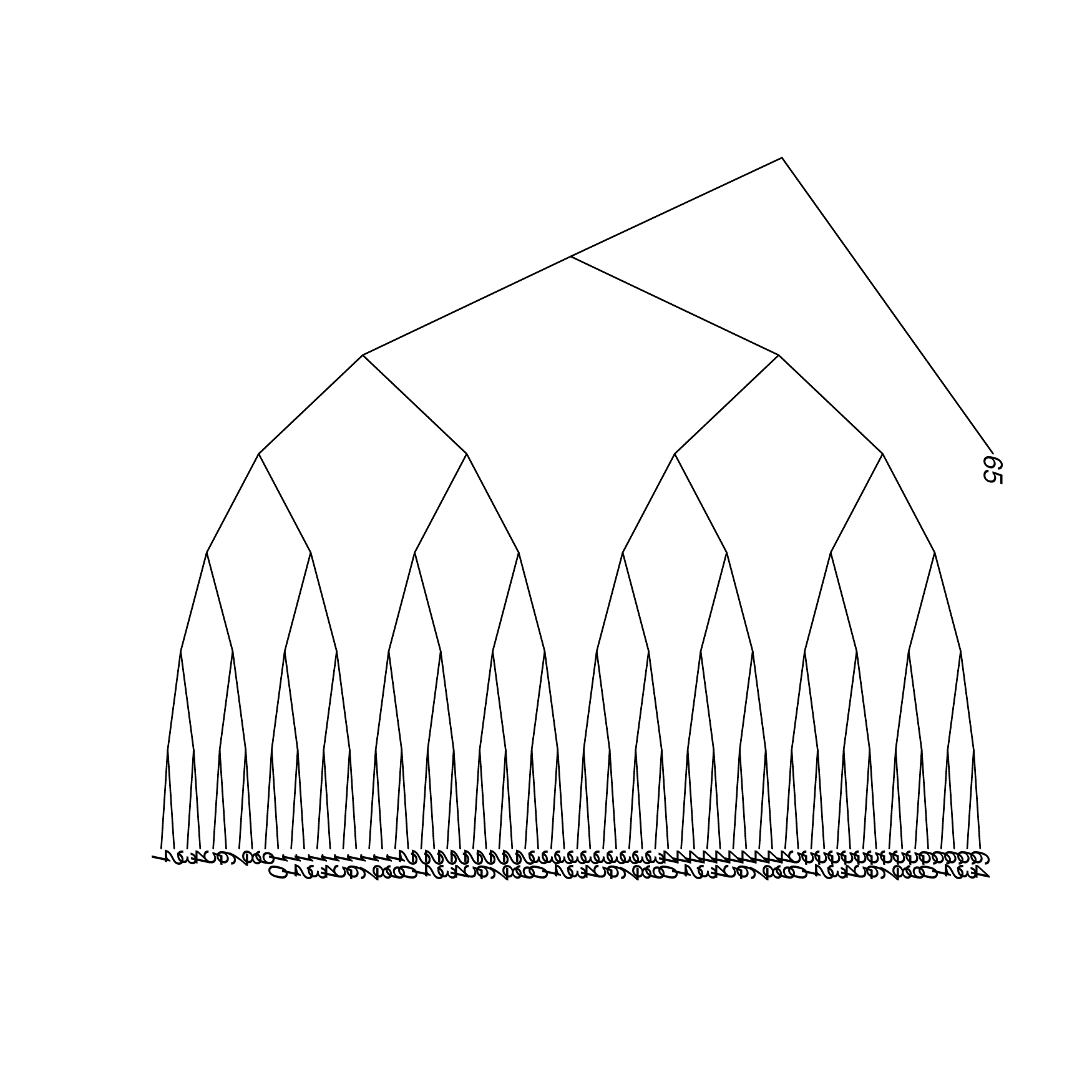}
\vskip-2cm
\caption{Balanced tree on 64 leaves, known as
the Bethe Lattice, we have added an outgroup to fix the root.}
\label{fig:tree64}
\end{figure}



\subsection{Seeing the Mutation Rate Gradient on Bethe Trees}
We generated 9 sets of trees with mutation rates set from $\alpha=0.01$ to
$\alpha=0.09$ and we generated the data
according to the Bethe lattice tree.\\
Here are the results in the first plane of the MDS:\\
\begin{figure}[h]
\vskip-1cm
\includegraphics[width=3.2in]{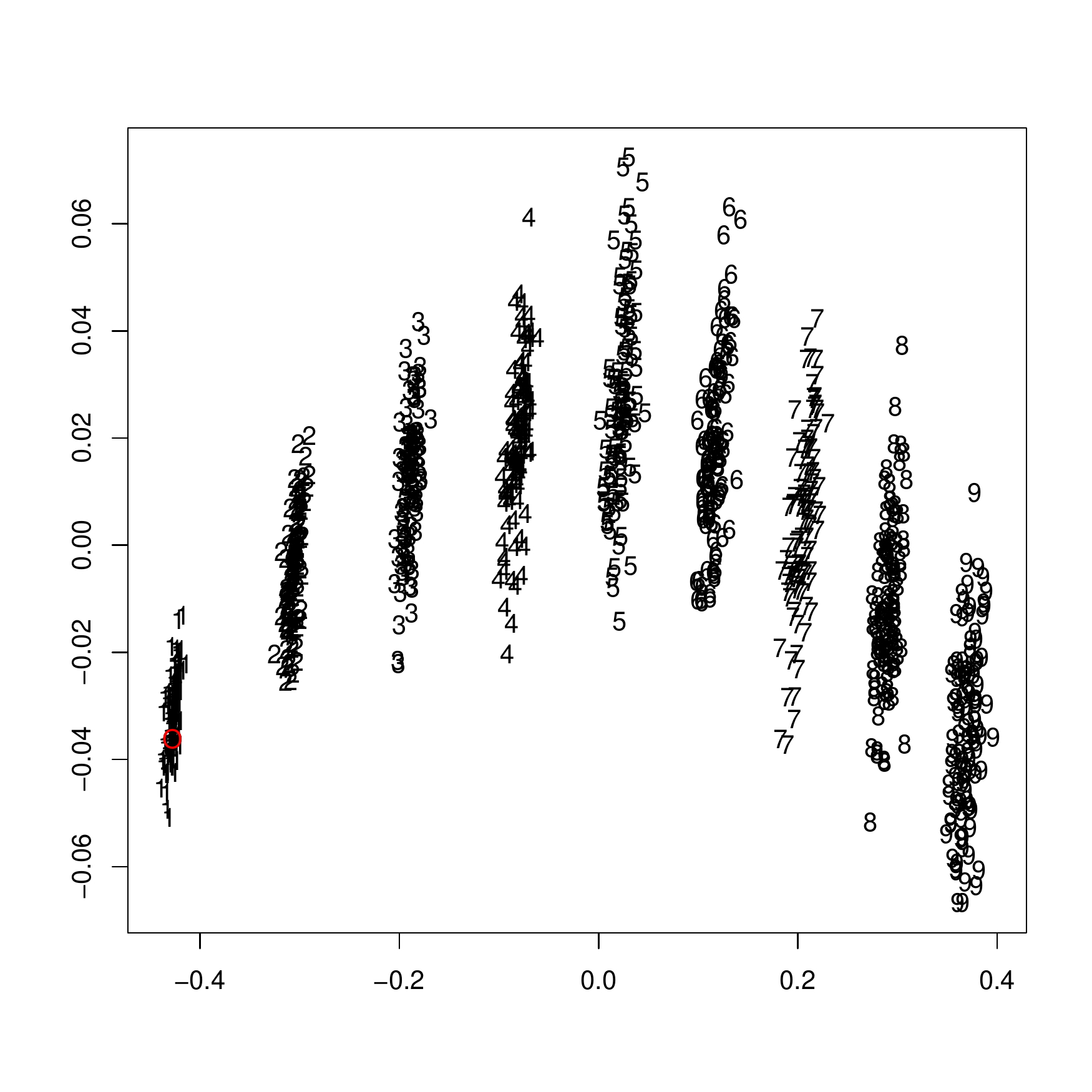}
\caption{The first two axes of the MDS of 901 trees with mutation rates
  varying from $\alpha=0.01$ to $\alpha=0.09$ (as labeled)}
\end{figure}
This shows that multidimensional scaling can be useful for comparing trees
which have differing mutation rates. The arch shape for the overall trend
is a classical instance of the horseshoe phenomenon \cite{goel}.  It is an
open question as to whether such a plot could be useful in the inverse
problem of trying to estimate the relevant mutation rate for a data set,
given the bootstrapped trees generated using the parametric bootstrap with
differing mutation rates.


\subsection{Finding inconsistent characters with high leverage}
In regression it is often useful to find observations with high
leverage. High leverage in regression can be detected by seeing a large
jump in the fitted model after the point is taken out. In the phylogenetic context, if
a character or a set of contiguous characters are taken out of the data
and the tree changes, this can be an indication of recombination or
horizontal transfer events.  In previous work \cite{Martins} use the the
minimum number of subtree prune-and-regraft (SPR) operations required to
resolve inconsistencies between two trees to detect recombination events
along DNA sequences in HIV.

Our approach is simpler since we will not use a Bayesian posterior, just
the distance between the original tree and the tree without the
character/segment.

Figure~ \ref{fig:hclustcv} in the first example show the MDS plot built
from distances between each of the cross-validation data sets built by
excluding a single gene and recomputing the hierarchical clustering tree
and then computing the {\bf BHV} distance between trees.  Each cross validated
tree is labeled by the gene that is excluded.  Table \ref{tab:distcv} in
the supplementary material shows the distances between the cross-validated
trees and the original tree.  If we consider the distances to the original
tree in the first row as shown in \ref{tab:col1}, we can see that they are
basically bimodal, a group around at a distance of about 17 from the
original tree and a group of values around 20 but the plot in
Figure~\ref{fig:hclustcv} tells a richer story since it shows how the
genes can be organized into three clusters according to the effect they
have on the overall hierarchical clustering tree. In some sense, this
gives us a picture of the leverage of each gene.
%


{\tiny
\begin{table}[ht]
\begin{center}
\begin{tabular}{rrrrrrrrrrrrrrrrrrr}
  \hline
 & Ori. & SNN & AF. & PLAG. & DHR. & PASK & PDE. &CAC.& LRR. & F2R &CX3.& MAD. & PPP. & KIF22 & MGC. & BCR & IFI. & TRIM.\\ 
  \hline
Ori. & 0 & 22 & 17 & 23 & 20 & 17 & 21 & 16 & 16 & 16 & 16 & 22 & 22 & 16 & 18 & 17 & 21 & 17 \\ 
    \hline
\end{tabular}
\end{center}
\caption{Rounded distances between the cross-validated trees and the original tree}
\label{tab:col1}
\end{table}
}

%

\section{Tree of trees}
\label{sec:treeoftrees}
A tree is a complete CAT(0) space   \citep{Gromov}.  Since \cite{Bhv} showed
that the space of trees is also a CAT(0) space (as shown in Figure~
\ref{fig:threetriangles}, this can be visualized by considering that
triangles are thin compared to Euclidean ones), we might guess a tree of
trees would be a satisfactory representation of a sample from the Bayesian
posterior or a bootstrap resampling distribution of trees.

We know that given a distance matrix we can give a treelike representation of the
points with these distances by building a tree if the distances obey  Buneman's four point
 condition  \citep{Buneman}.\\
\specialbox{Buneman's four point condition}{
For any four points 
$(u,v,w,x):$\\
The two largest of the three sums:$
d(u,v)+d(w,x),
d(u,w)+d(v,x),
d(u,x)+d(v,w)$
are equal.
}

We can see
Gromov's  definition the hyperbolicity constant $\delta$
  \citep{Gromov} as a relaxation of the above four-point condition:\\
\specialbox{ Gromov's   hyperbolicity constant}{
For any four points u,v,w,x, the two larger of the three sums $d(u,v)+d(w,x),
d(u,w)+d(v,x),
d(u,x)+d(v,w)$
differ by at most 2$\delta$. 
}

 We propose using the smallest $\delta$ that works
as a numeric criteria to quantify how treelike
a set of finite points with a geodesic metric is.

Conside the  bootstrap example in section 3.3, where we have
$B\times B$ matrix of distances. We can use the following to compute $\delta$:\\
\specialbox{
Algorithm to compute the $\delta$-hyperbolicity}{\tt
For all sets of 4 points among $B$, call them (i,j,k,l) \\
(of which there are $R=\frac{B(B-1)(B-3)(B-4)}{24}$).\\
 \mbox{For }( r\mbox{ from }1\mbox{ to }R) do (1)(2)(3) below:
 \begin{enumerate}
\item
Compute $
A_1=d_{ij}+d_{kl},\qquad
A_2=d_{ik}+d_{jl}\qquad
A_3=d_{il}+d_{jk}.$
\item
Sort
$A_1,A_2, A_3 \mbox{ giving } (A_{(1)},A_{(2)}, A_{(3)})\mbox{ in decreasing order.}$
\item
$\mbox{Take }E(r)_=(A_{(1)}-A_{(2)})$
\end{enumerate}
$\mbox{Take }\delta=\frac{1}{2}\max_{r} E(r)$
}

The algorithm is implemented in the \texttt{distory} package. It takes
about a minute to compute the $\delta$ -hyperbolicity of a $500\times 500$
distance matrix.  In order to calibrate how small delta becomes both on
uniformly distributed data in Euclidean space and on a finite treelike
data set generated from balanced Bethe trees, we ran simulations where we
generated points from known trees and then computed various distances
between them.

\begin{table}[ht]
\label{tab:delta}
\begin{center}
\begin{tabular}{|llrllll|}
  \hline
  Data & Distribution & Distance& Max (sd)& Mean(sd) & $\delta$ (sd)& $\delta$/Max (sd)\\
\hline
\hline
500 points  &Uniform& Manhattan
&
13.8 (0.33) & 8.33 (0.04) &7.03 (0.26)& 0.51 (0.02)\\
500 points & Uniform & Euclidean& 
3.04 (0.06) & 2.03 (0.009) & 1.38 (0.05) &0.45 (0.02)\\
\hline
512 points & MVNormal & Manhattan&
49.14 (1.59) & 28.22 (0.20) &21.45 (0.79) &
0.44 (0.02)
\\
512 points & MVNormal & Euclidean &
11.66 (0.41)   & 7.00 (0.05)   & 4.82 (0.17)  & 
0.41 (0.02)
\\
\hline
512 points &Bethe Tree&
      JC69 &  
%
  0.223 (0.008) & 0.16 (0.003) &0.017 (0.001)&
                    0.076
       (0.0043)
             \\
            512 points &Bethe Tree&
      Raw & 
      0.19 (0.006) &  0.14 (0.002) &  0.013 (0.001) &
0.069 (0.004)
 \\
\hline     
\hline
500 trees & \texttt{rtree} & BHV & 8.03& 6.46 & 0.68 &0.085\\
30 leaves & uniform splits & & & & &\\
\hline
500 trees & \texttt{rtree} & BHV & 8.94 &7.53 & 0.60& 0.067\\
40 leaves & uniform splits & & & & &\\
\hline
500 trees & \texttt{rtree} & BHV &9.91 & 8.52 &0.63 & 0.064 \\
50 leaves & uniform splits & & & & & \\
\hline
500 trees & \texttt{rtree} & BHV & 10.45& 9.32 &0.59  & 0.056\\
60 leaves & uniform splits & & & & &\\
\hline
500 trees & \texttt{rtree} & BHV & 11.50& 10.09& 0.59&0.051\\
70 leaves & uniform splits & & & &  &\\
\hline
\end{tabular}
\caption{Different values of $\delta$ and the ratio $\delta$/max(d) for
  points generated both in bounded Euclidean space and for points
  generated from trees. In the top half of the table, each value was estimated from 100 simulations, in
  the Euclidean case the distances were computed from points generated in
  25 dimensions. In the lower half of the table the values were generated by 
  randomly generated trees of a given size, 30,40,50,60 and 70 leaves and looking at the maximum
  and $\delta$ values for this sample of 500 trees. We can see that the sets of randomly generated trees have a much lower
  $\delta$-hyperbolicity and little variation with the number of leaves in the tree.}
\end{center}
\end{table}

In particular, we used the $\delta/max$ statistic in the case of the
bootstrapped trees represented by the MDS plot in the resulting ratio was
$0.47$. This indicates given, the calibration experiments in the above
table, that point configuration would be well approximated by a Euclidean
MDS. The $\delta /max$ statistic is a rough approximation for scaling each
triangle considered by its diameter; two other approximations, scaling by
the perimeter and scaling by the max of the sums $A_{(1)}$ are implemented
in the R package.

We note here that recently, theoretical computer scientists have
used the geometry of graphs for algorithmic purposes
 following the breakthrough paper by \cite{Linial}.
 The computation of $\delta$-hyperbolicity provided here could also be used in this context.
 There is also a healthy literature connected to the subject of metric graphs that includes other applications
 of $\delta$ hyperbolicity,
 for a  comprehensive review see 
 \cite{BandeltChepoi}.


\subsection{Mixture Detection}
A particularly interesting application of this idea is in the detection of
mixtures of the evolutionary processes underlying aligned sequences.
Mixtures pose problems when using MCMC methods in the Bayesian estimation
context \citep{MosselV}. These authors note that MCMC methods in particular
those used to compute Bayesian posterior distributions on trees can be
misleading when the data are generated from a mixture of trees, because in
the case of a `well-balanced' mixture the algorithms are not guaranteed to
converge. They recommend separating the sequences according to coherent
evolutionary processes.  But this means the first step in the analyses of
aligned sequences should be the detection of the mixture and proposals for
separating the data at certain positions.

Suppose the data come from the mixture of several different trees; we will
see how the bootstrap and the various distances and representations can
detect these situations.

Our procedure uses the bootstrap or a Bayesian posterior distribution.
Suppose we have $\tau_1,\tau_2,\ldots, \tau_K$ $K$ trees generated from
one original alignment either by bootstrapping the original data or by
using a MCMC method for generating from the Bayesian posterior.  We use
the distance between trees to make a hierarchical clustering tree using
single linkage to provide a picture of the relationships between the
trees.

In this simulated example we generate two sets of data of length $1,000$
from the two different trees.
We concatenate the data into one data set on which the standard
phylogenetic estimation procedures are run.  This provides the estimated
tree for the data. We then generate 250 bootstrap resamples from the
combined data, and compute the distances between the 250 trees from each
of the bootstrap resamples, using these to make a hierarchical clustering
single linkage from this distance matrix. \\

\begin{figure}[ht] 
\vskip-1cm
\includegraphics[width=5in]{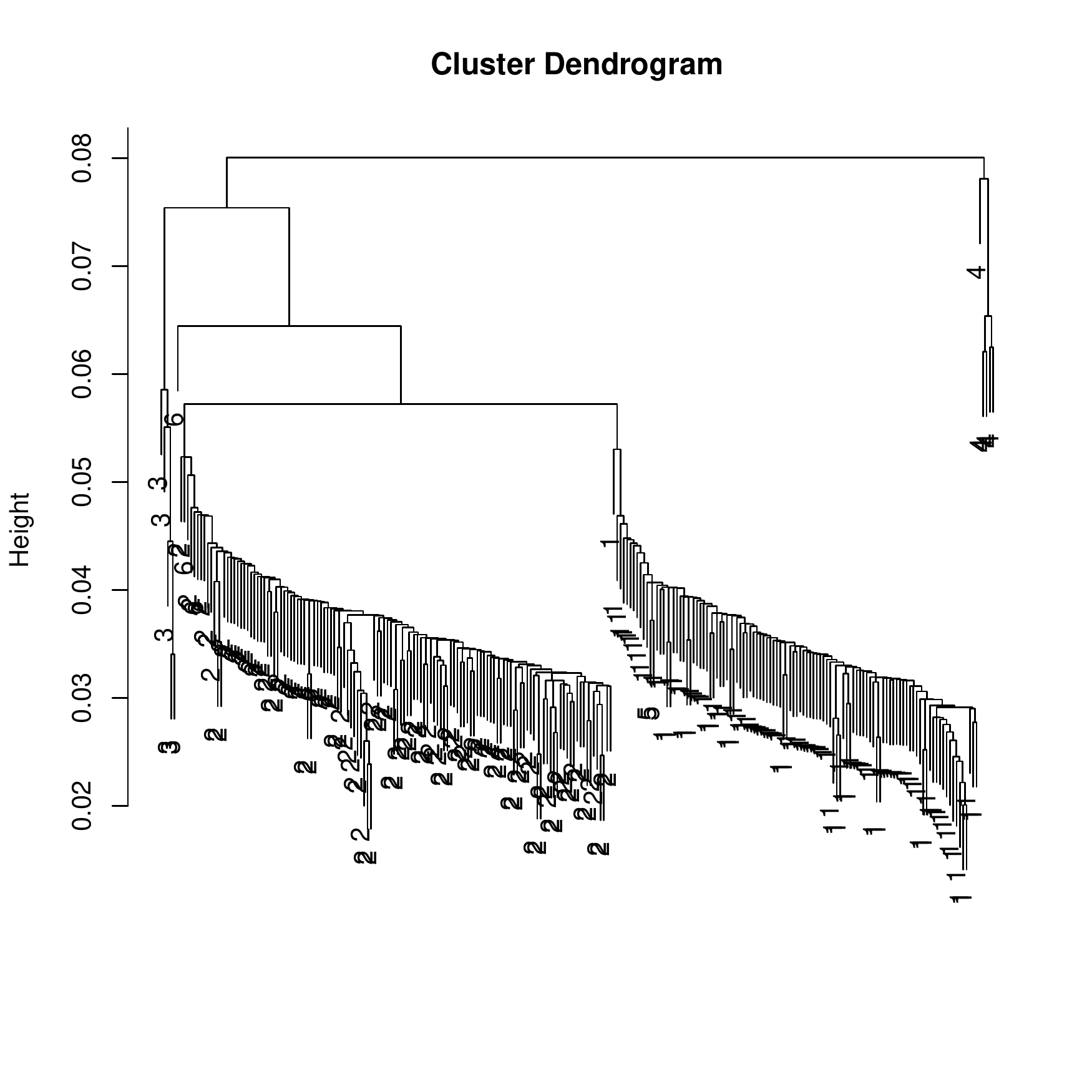} 
\vskip-2cm
   \caption{Hierarchical clustering of 250 trees resulting from a nonparametric bootstrap
   of the data generated by the double data set $\mathcal{X}_{12}$}
   \label{fig:example}
\end{figure}

\subsection{Variability of Trees from a Bayesian Posterior Distribution}
After running MCMC Bayesian sampling from the posterior such as that
available in \texttt{MrBayes}\cite{MrBayes} we obtain several sets of
trees from different runs of the chain. In order to evaluate these picks,
we took 250 random picks from the two runs combined, with the first
200,000 trees from each run discarded. Each MCMC was run 1,000,000 times
on the same subset of the Laurasiatherian data available in the
\texttt{phangorn} package  \citep{phangorn}.  We computed the $\delta/max$
statistic for the distance matrix between all 250 trees and obtained a
value of $0.57$ indicating that the trees could be well approximated by a
Euclidean representation.

\begin{figure}[h!] 
   \centering
   \includegraphics[width=4in]{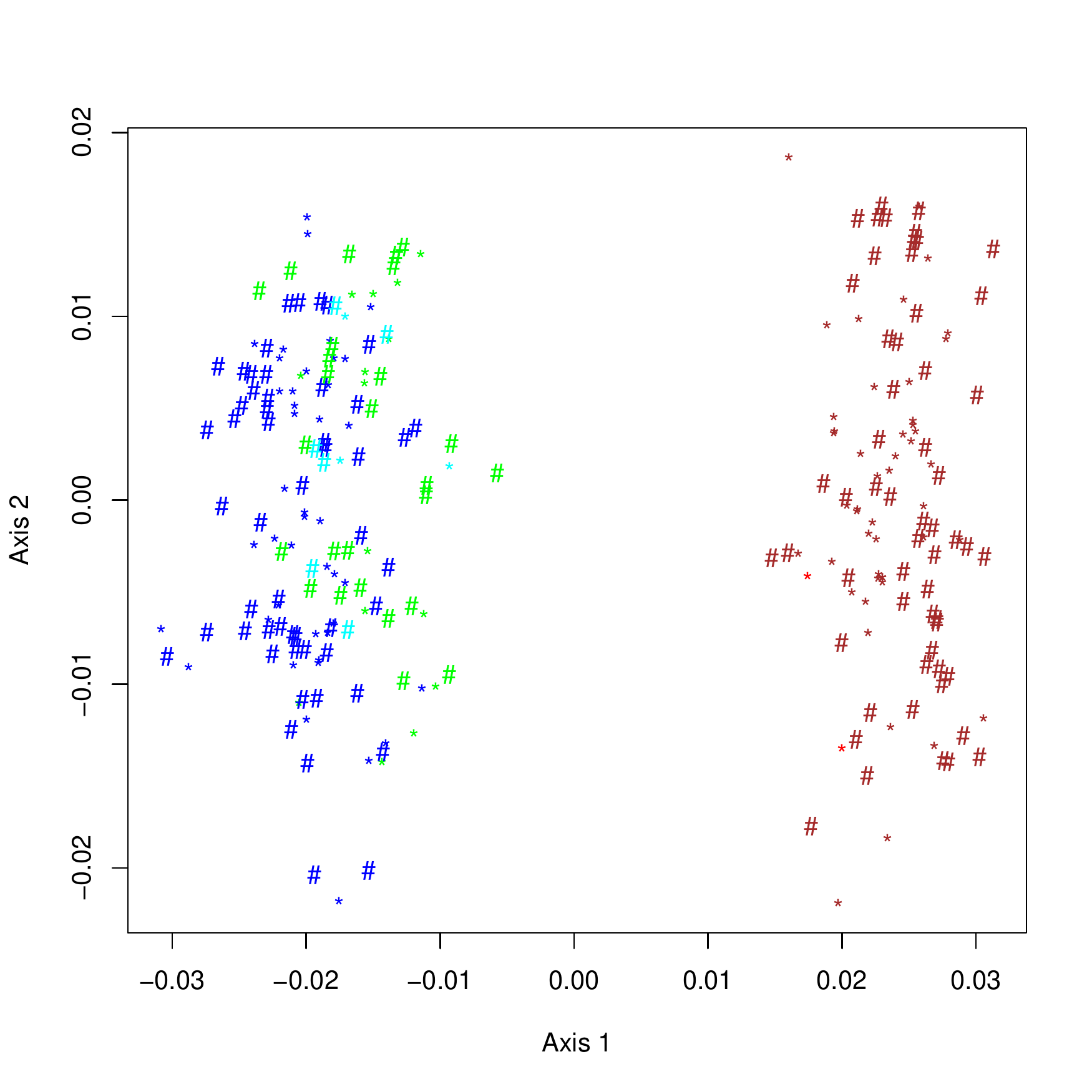}
   \caption{First plane of MDS plot (41\%) of the trees sampled from the
     Bayesian posterior. The cloud on the right contains trees all from
     the same branching pattern (color indicates branching pattern), with
     an equal number of picks from the first run as from the second.}
   \label{fig:bayeslaura}
\end{figure}

The standard MDS plot is shown in Figure~\ref{fig:bayeslaura}. We see that
the scatterplot is bimodal, but this cannot be explained either by the
runs, the trees sampled during the first run are given as '*' whereas the
trees from the second runs are the "\#" character.

There were 5 different branching patterns in all, from run one there $(42
, 2 ,15 ,47 , 3)$ of each of the five categories of trees and $( 52,0, 27,
57, 6)$ in the second run (note both tuples represent counts of the same 5
topologies).  We have colored each of these with one of five colors.

%
\subsection{Kernel Multidimensional Scaling}
It often makes sense to transform the distances 
so that the graphical representation focuses on correctly
representing 
differences between similar points and doesn't try to
make a good representation of the distances between points which are far
apart. This has been shown to be particularly effective when the original
points lie on a postiviely curved manifold as for instance in
  \citep{Tenenbaum:2000,Roweis:2000}. However it can also be useful in the context of trees where
the very local distances are Euclidean and the negative curvature only appears as points get
further apart. Thus we will use the exponential kernel that takes the
 distance between trees $d(T_1,T_2)$  and compute
a new dissimilarity measure between trees as $\delta(T_1,T_2)=1-exp(-\lambda d(T_1,T_2))$
Then $\delta(T_1,T_2) \approx \lambda d(T_1, T_2)$ for $d(T_1, T_2) \ll 1$ and 
$\delta(T_1,T_2)\approx 1$ if $d(T_1,T_2)$ is large, so it will not be
 sensitive to noise around relatively large values of $d(T_1, T_2)$.

This kernel has the effect of representing carefully relations between close trees and thus is very useful for exploring local neighborhoods of treespace.

\subsection{Learning the Right Distances}
There are reasons to think that
these distances are not
giving us all the information we are looking for. 
If this is the case, we could ask the inverse question, given a set of trees and a distance
which seems to be a meaningful representation of similarities, we can ask the inverse
question of how to combine the topological information about the tree into the `right distance'. 
For instance if we are told that  set $S_1$ of trees should be similar
and different from another set $S_2$, we can make a parametric model
that weights partitions on the tree so as to be concordant with this notion.
There have been similar approaches to problems in text analysis solved by what is
often called ``metric learning", for a recent illuminating example see \cite{Lebanon}.

In multivariate analysis it is often important to account for differing levels of variability in the data
by rescaling variables. In the context of phylogenetic trees,
for instance for the case where the contemporary DNA sequences are used to build trees
that go far back in the past,
it seems  natural to ask if it wouldn't be better to put differential weights on the branches of the 
tree to compensate for the higher uncertainty with which we can only infer what is happening high up
in the tree\citep{Mossel:2004}. In the same way we rescale variables so they have the same variance before doing
a multivariate analysis, we would divide the edges in the tree closer to the root with larger numbers corresponding to the larger uncertainty, so that large differences higher in the tree would be downweighted as we are not sure of them.
Examples of this differential weighting can be found in the \texttt{vignette} that accompanies the
package \texttt{distory}   \citep{distory}.

\section{Using the path between two trees to find boundaries}
It can be useful to explore both the neighborhood of a given tree and the
datasets which are borderline in the sense that small perturbations induce
a change in the tree topology.

\subsection{Paths between different tree topologies}
How close our estimate is to being {\em borderline}, in a particular sense
of closeness, will inform us on the stability of our estimate.  This can
be done by creating small perturbations of the original data by
bootstrapping. Looking at the set of bootstrapped trees, we can study the
changes possible from small perturbations induced on the tree, both as
changes in the topology and as quantitative measurements of distance.  If
all the bootstrap resamples give the same tree, then we are sure that the
estimate we have is not "borderline," that is, the topology of the
estimate is the same as that of the true tree. 

However if the bootstrap resamples give many different trees it may be
that the original data are not very treelike and the inferred tree has
many competing neighbors.
\begin{figure}[h]
\includegraphics[width=3.2in]{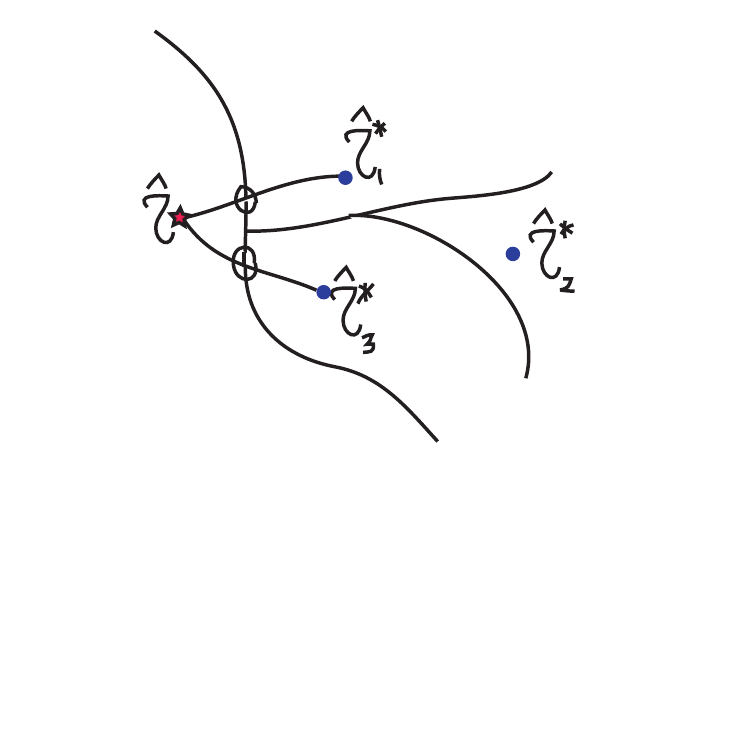}\\
\vskip-3cm
\caption{Example of three bootstrap trees (with stars) and the tree
  estimated from the original data.  We see that trees {\Large
    $\hat{\tau}^*_1$} and {\Large $\hat{\tau}^*_3$} share a common
  boundary with the original estimate {\Large $\hat{\tau}$}.  We are
  particularly interested in {\em borderline trees}, ie trees at the
  intersection between the paths between trees and the boundaries between
  regions defined by each topology.}
\end{figure}

It can be interesting to know whether all the alternative bootstrap trees
are the same, as happens in the case of an underlying mixture of trees, or
whether there are a large number of competing trees, this will incline us
more towards concluding that the data are far from treelike. In fact, if
the star tree with all edges equal to zero is close to the original tree
then the number of alternatives will be exponentially large   \citep{ihp}.
If $r$ contiguous edges of the tree are small, there will still be
$(2r-3)!!=(2r-3)\times(2r-5)\times..3\times 1$ trees in its close
neighborhood.

In the case of the bootstrap analysis whose binned topologies are
presented in Table \ref{tab:binLaura} we found that there are actually 9
trees in the bootstrap resample of that are borderline neighbors to the
original tree, these neighbors and their respective BHV distances to the
original tree are presented in the supplementary data (Figure~
\ref{fig:Neigbors9}).  We see that there are thus many `small edges' in
this particular tree estimate, and the original estimated tree shares
boundaries with $9$ competing orthants.
\subsection{Finding Borderline Data with MCMC}
After we estimate a tree based on a specific set of DNA sequences, we may
be interested in seeing which particular changes to the
original sequences may lead to alternative topologies. 

We can use Markov Chain Monte Carlo methods to help with this: given a
target tree with the topology of interest, we try to find a configuration
of weights for the columns from the original sequences such that the tree
estimated from data sampled from those weights is as close to the target
tree (on the boundary) as possible.

We start with the original sequences, such that every position in the
sequences occurs once; we then draw proposals of increasing the count of
one position by one and decreasing the count of another by one, to
maintain the same number of total base pairs in the dataset (intuitively,
we write over the DNA at one position with that from another position in
such a way that the DNA at the position we wrote over can still be used
again). A tree is estimated with the proposed change, and the proposal is
accepted or rejected based on the ratio of the old distance to the new
distance (if the new distance is smaller, the ratio is greater than 1, so
the proposal is accepted automatically; if the new distance is greater,
with some probability the proposal is accepted anyway to allow the MCMC to
get out of local minima).  A simulated annealing scheme
  \citep{Kirkpatrick} introduces a temperature which is used to help with
location of a closest approximating set of positions, which are then
reported along with the resulting tree and the final distance.  This
algorithm has been implemented in \texttt{R}   \citep{R} and is also
available in the \texttt{distory}   \citep{distory} package.

\section{Conclusions : More questions than answers.}
We have combined the problems of evaluating phylogenetic trees and
hierarchical clustering displays in a common mathematical framework. By
embedding rooted binary trees in a metric space associated to the BHV
distance, we can capitalize on statistical methods such as MDS to solve
some of the difficulties in evaluating distributions of trees as output by
Bayesian posterior sampling or bootstrap methods.

We show through examples that a distance between trees can provide
valuable information about the variability of tree estimates and a
substitute notion of multivariate spread. However, many questions remain
unanswered. 

At a fundamental level, what are realistic probability measures on
treespace and how they should be used to provide useful priors and a
theoretical notions of variance and more general moments on the space? 

More generally, we face the issue of a practical way of quantifying
variability of points. This question in tree space has a resemblance with
phylogenetic diversity   \citep{Faith}. Trees could be considered leaves on
a tree of trees, and thus a more nuanced measure of diversity than a
simple Shannon type index is available.

Here we propose Gromov's $\delta$-hyperbolicity constant as a useful
statistic for evaluating if the points in treespace were better
approximated by a tree of trees or a Euclidean approximation.  However,
many open questions remain around this topic. We have only touched briefly
on the question of the correct scaling of the $\delta$ statistic; our
choice was to take the maximum distance between two points, but other
choices can be argued, recently \cite{Bonahon} choose a more `exact'
normalization that  comes with high computational price, their suggestion 
is to
compute the diameter of all the geodesic triangles (two simpler methods,
using the perimeter and max of the sums, may be good approximations as
well and are available in the \texttt{distory} \citep{distory} package).

Nothing is known about the distributional theory of Gromov's $\delta$
statistic. Its dependence on the underlying dimension and sparsity of the
data is obvious, but there is much more work to be done before valid
inferential methods become available for evaluating both the local and the
global curvature of a finite set of points.

Finally we have shown how the BHV distance can be used to evaluate the
distance to the boundary tree closest to it and to some of the data sets
that could have given such a tree. More work needs to be done to provide accurate leverage statistics for trees.

\section*{Acknowledgements}
We thank Persi Diaconis for a careful reading of an earlier version.

\bibliographystyle{plainnat}
\bibliography{disTree}
\clearpage
\section*{Supplementary Data}

{\tiny
\begin{table}[ht]
\begin{center}
\begin{tabular}{rrrrrrrrrrrrrrrrrrr}
  \hline
 & Ori. & SNN & AF. & PLAG. & DHR. & PASK & PDE. &CAC.& LRR. & F2R &CX3.& MAD. & PPP. & KIF22 & MGC. & BCR & IFI. & TRIM.\\ 
  \hline
Original & 0 & 22 & 17 & 23 & 20 & 17 & 21 & 16 & 16 & 16 & 16 & 22 & 22 & 16 & 18 & 17 & 21 & 17 \\ 
  SNN & 22 & 0 & 33 & 34 & 32 & 34 & 32 & 29 & 32 & 30 & 33 & 31 & 34 & 33 & 33 & 33 & 33 & 33 \\ 
  AF5q31 & 17 & 33 & 0 & 34 & 31 & 17 & 31 & 18 & 16 & 17 & 15 & 31 & 34 & 15 & 16 & 12 & 32 & 15 \\ 
  PLAG1 & 23 & 34 & 34 & 0 & 33 & 29 & 14 & 32 & 31 & 32 & 30 & 18 & 10 & 31 & 32 & 34 & 31 & 33 \\ 
  DHRS3 & 20 & 32 & 31 & 33 & 0 & 31 & 30 & 27 & 29 & 26 & 29 & 31 & 33 & 28 & 32 & 29 & 25 & 31 \\ 
  PASK & 17 & 34 & 17 & 29 & 31 & 0 & 30 & 18 & 12 & 17 & 9 & 33 & 30 & 12 & 13 & 16 & 28 & 14 \\ 
  PDE9A & 21 & 32 & 31 & 14 & 30 & 30 & 0 & 31 & 29 & 31 & 30 & 12 & 11 & 29 & 31 & 33 & 30 & 33 \\ 
  CACNB3 & 16 & 29 & 18 & 32 & 27 & 18 & 31 & 0 & 14 & 9 & 17 & 30 & 32 & 17 & 18 & 14 & 30 & 18 \\ 
  LRRC5 & 16 & 32 & 16 & 31 & 29 & 12 & 29 & 14 & 0 & 15 & 10 & 32 & 31 & 9 & 11 & 14 & 27 & 16 \\ 
  F2R & 16 & 30 & 17 & 32 & 26 & 17 & 31 & 9 & 15 & 0 & 16 & 31 & 33 & 15 & 18 & 12 & 31 & 17 \\ 
  CX3CR1 & 16 & 33 & 15 & 30 & 29 & 9 & 30 & 17 & 10 & 16 & 0 & 33 & 31 & 8 & 11 & 12 & 27 & 16 \\ 
  MAD-3 & 22 & 31 & 31 & 18 & 31 & 33 & 12 & 30 & 32 & 31 & 33 & 0 & 16 & 32 & 34 & 32 & 33 & 32 \\ 
  PPP2R2B & 22 & 34 & 34 & 10 & 33 & 30 & 11 & 32 & 31 & 33 & 31 & 16 & 0 & 32 & 32 & 35 & 32 & 33 \\ 
  KIF22 & 16 & 33 & 15 & 31 & 28 & 12 & 29 & 17 & 9 & 15 & 8 & 32 & 32 & 0 & 11 & 14 & 27 & 17 \\ 
  MGC4170 & 18 & 33 & 16 & 32 & 32 & 13 & 31 & 18 & 11 & 18 & 11 & 34 & 32 & 11 & 0 & 16 & 31 & 18 \\ 
  BCR & 17 & 33 & 12 & 34 & 29 & 16 & 33 & 14 & 14 & 12 & 12 & 32 & 35 & 14 & 16 & 0 & 31 & 14 \\ 
  IFI16 & 21 & 33 & 32 & 31 & 25 & 28 & 30 & 30 & 27 & 31 & 27 & 33 & 32 & 27 & 31 & 31 & 0 & 32 \\ 
  TRIM26 & 17 & 33 & 15 & 33 & 31 & 14 & 33 & 18 & 16 & 17 & 16 & 32 & 33 & 17 & 18 & 14 & 32 & 0 \\ 
   \hline
\end{tabular}
\end{center}
\caption{Rounded distances between cross validated hierarchical clustering  trees and the one computed using all the genes.}
\label{tab:distcv}
\end{table}
}

\begin{figure}[ht!] 
\begin{center}
     \includegraphics[width=7.5in]{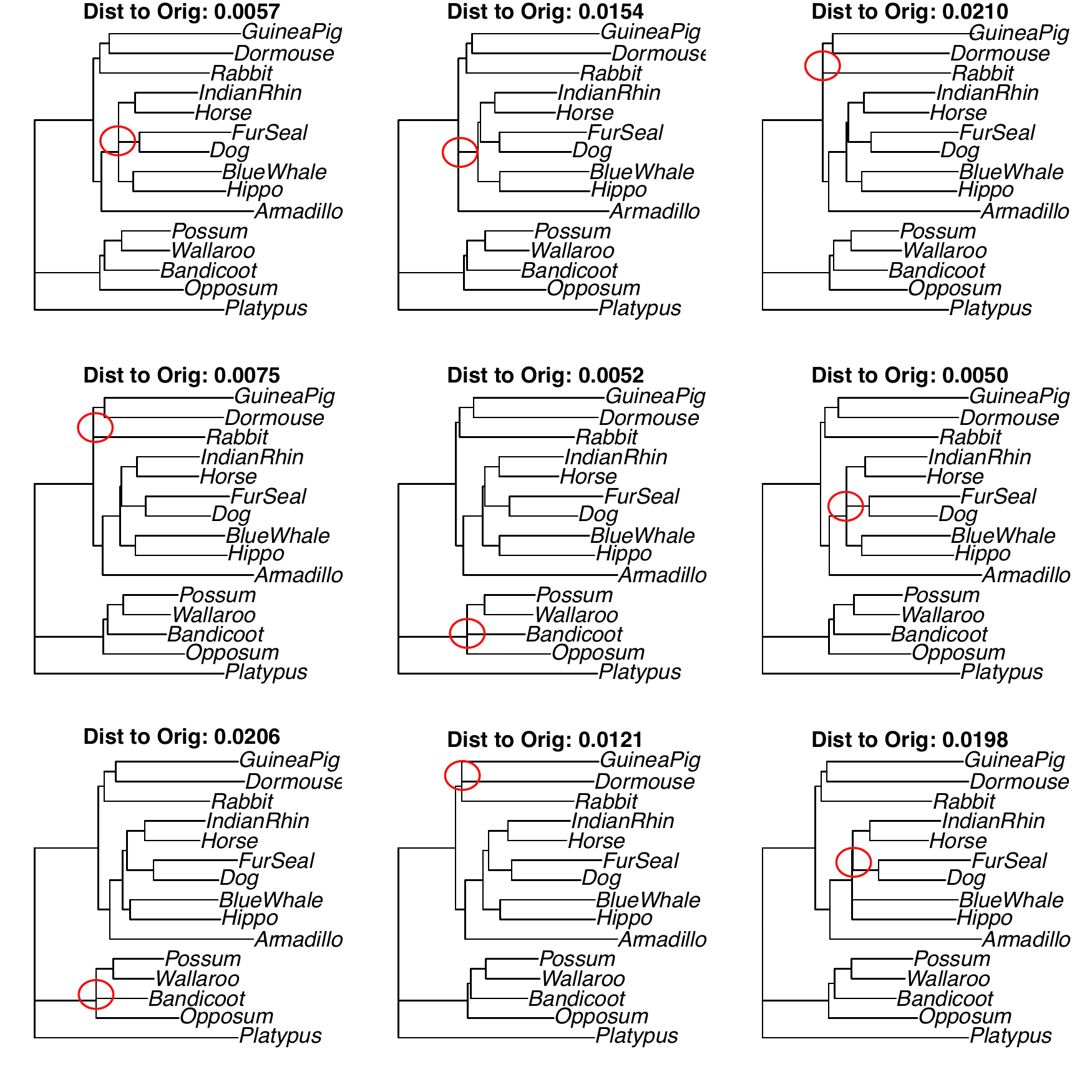} 
     \caption{All 9 neighboring trees from the bootstrap of the Laura12 tree.}
   \label{fig:Neigbors9}
   \end{center}
\end{figure}
\end{document}